\documentclass[]{aa}  
\usepackage{algpseudocode}

\usepackage{graphicx}
\usepackage{txfonts}

\begin{document} 
\authorrunning{Ferrone et al}
\titlerunning{IMB ancient dynamical family \& planetesimals}

   \title{Identification of a 4.3 billion year old asteroid family and planetesimal population in the Inner Main Belt}

   \author{S. Ferrone\inst{1,2}, M. Delbo\inst{1}, C. Avdellidou\inst{1}, R. Melikyan\inst{1,3,5}, A. Morbidelli\inst{1}, K. Walsh\inst{4}, \and R. Deienno\inst{4} }

   \institute{Observatoire de la Côte d'Azur, CNRS-Lagrange, Université Côte d'Azur, CS 34229, 06304 Nice Cedex 4, France.
         \and
             LESIA \& GEPI, Université Paris Cité, Observatoire de Paris, Université PSL, Sorbonne Université, CNRS, F-92190 MEUDON, France.
          \email{salvatore.ferrone@obspm.fr}
         \and 
            Lunar and Planetary Laboratory, University of Arizona,1629 E University Blvd, Tucson, AZ 85721, USA.
         \and 
            Southwest Research Institute, 1050 Walnut St. Suite 300, Boulder, CO, 80302, USA.
            \and
            Department of Physics and Astronomy, Ithaca college, 953 Danby Road, Ithaca, NY 14850, USA. }

   \date{Received 01 Dec 2022 / Accepted 22 May 2023}

 
  \abstract
   {Understanding the conditions that lead to the formation of planetesimals---the building blocks of planets---and their initial size distribution is a central problem of modern planetology. While most of these original planetesimals were accreted onto the terrestrial planets and the cores of the giant planets, some were also stranded in the main belt, where 4.5 Gyr of collisional evolution broke most of them into families of collisional asteroid fragments. However, some planetesimals survived, and are still hidden amongst asteroid fragments in the main belt. }
   {We make use of astronomical data to identify these leftover planetesimals amongst all other asteroids. Our search is based on separating planetesimal survivors from families of asteroids generated by collisions. Namely, we aim to identify and “clean” the main belt of collisional family members: by doing so, we would be left with the surviving members of the original planetesimals.}
   {We focus here on the inner portion of the main belt for asteroids with intermediate to high albedo. It is known that current asteroid family catalogs are not suitable for the aforementioned cleaning; they are conservative and only one-quarter of the known asteroids are associated with the approximately 120 distinct asteroid families. We therefore developed methods to inclusively link asteroids to known collisional families in order to better capture their extent. Namely, we apply a hierarchical clustering method (HCM) on asteroids filtered according to the V-shape of the Yarkovsky drift of  each family in order to reassess family membership (V-shape-constrained HCM). The identified families were removed and the remaining background population was searched for previously undetected collisional families. }
   {We succeed in using our V-shape-constrained HCM to link family ``halos'' to their cores. After removing these reassessed families from the asteroid population, our V-shape search reveals a previously unknown collisional family of S-type asteroids in the inner main belt with an age of $4.3\pm1.7$ Gyr and a significance level of 3.4$\sigma$. When this ancient collisional family is removed, 34 planetesimals are identified and their size--frequency distribution is presented.}
{The asteroid belt has two components: planetesimals and collisional fragments. The cumulative size--frequency distribution of planetesimals has a steep power-law index for bodies larger than 100 km in diameter and a much smaller power-law index for planetesimals smaller than 100 km. }

 
\maketitle
%

\section{Introduction}
\label{S:intro}

Understanding the formation of planets and small bodies is a central problem of planetary science. It is recognized that the first stage of this process is the accretion of the so-called planetesimals from the solids in protoplanetary disks \citep[e.g.,][and references therein]{Birnstiel2016SSRv..205...41B}. Studies of the formation of planetesimals have  made tremendous progress in the last few decades, and the consensus is that these objects formed from clumps of solid particles that reached sufficient densities to become self-gravitating and to contract to form asteroid-sized bodies \citep{Johansen2007Natur.448.1022J}, a process also favored by disk turbulence \citep{Johansen2007Natur.448.1022J,cuzzi2008toward} and the Kelvin-Helmholtz instability \citep{Johansen2007Natur.448.1022J}. However, most studies approach the problem of planetesimal formation theoretically, as only limited constraints  are available on the original planetesimal size and composition distribution in our Solar System \citep[e.g.,][and references therein]{klahr2022}.

\begin{figure*}[t!]
    \centering
    \includegraphics[width=\textwidth]{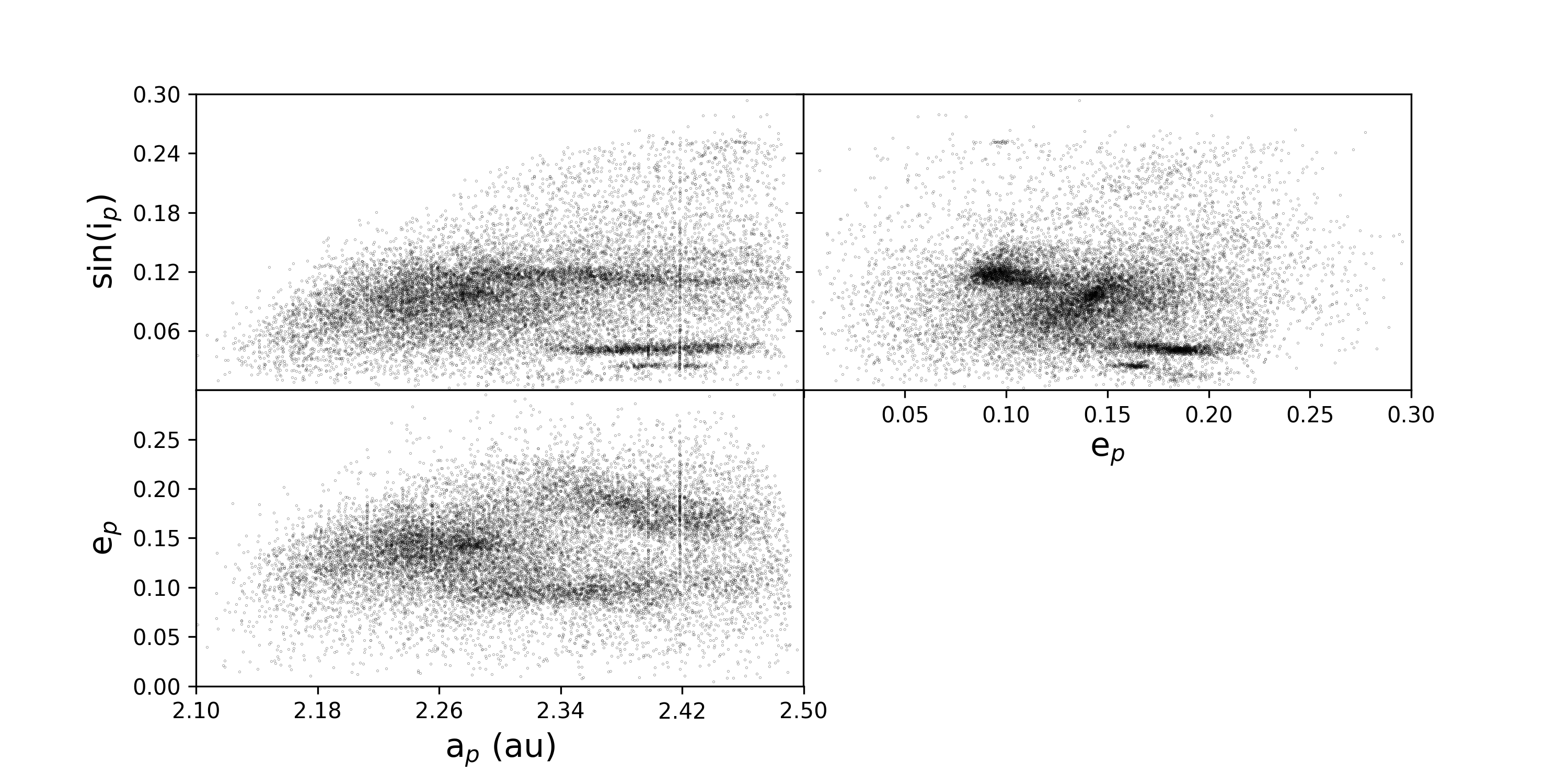}
    \caption{Proper orbital elements of the inner main asteroid belt. Only bodies with $p_{V}$ greater than 0.123 are shown. The y-axis is shared for the top two panels, while the x-axis is shared for the left two panels. Clusters of objects correspond to some of the most preeminent asteroid families.}
    \label{fig:backgroundHighAlbedo}
\end{figure*}
    
In order to obtain further such constraints, significant effort has been made to investigate the population of the minor bodies of our Solar System \citep{Bottke2005Icar..175..111B,Morbidelli2009Icar..204..558M,Delbo2017,tsirvoulis2018reconstructing}. Indeed, asteroids and all the small bodies, such as comets, trans-Neptunian objects, and irregular satellites, are what is left of the original planetesimal disk from the planet-formation era. However, not all asteroids are survivors from primordial times: a large number of these are collisional fragments of the original planetesimals \citep{Delbo2017,tsirvoulis2018reconstructing,Dermott2018NatAs...2..549D}. Although these fragments still carry the original composition of their progenitors, their sizes, spin vector orientations, and shapes, they do not provide direct information about the accretion processes that led to the formation of planetesimals and, consequently, of the planets.
    
With this  in mind, different studies have attempted to separate the original asteroids that accreted as planetesimals in the protoplanetary disk from the collisional fragments \citep{Delbo2017,Delbo2019,tsirvoulis2018reconstructing,de2022composition}. These works are based on the concept of identifying all families of fragment asteroids that were produced in the main belt after the early formation phases of our Solar System and “cleaning” them from the main belt in order to highlight the survivors of the original asteroids, that is, the planetesimals.

The question remains as to how we can best distinguish asteroid fragments from planetesimals. In a break up process, fragments are launched into space at moderate velocities (some m~s$^{-1}$), which is how, in the main belt, fragments have similar orbital elements to those of their parent body \citep[e.g.,][and references therein]{Vokrouhlicky2006}. Thus, fragments become new asteroids themselves, clustered in orbital space, as shown for example in Fig.~\ref{fig:backgroundHighAlbedo}. Additionally, asteroids inside these families have,
in general, similar physical properties, such as geometric visible albedo ($p_V$) \citep{masiero2015asteroid}, color, and spectrum \citep{Parker2008Icar..198..138P,collaboration2022gaia}. These clusters of fragments are the so-called asteroid families and are typically identified using the yierarchical clustering method (HCM). This method looks for clusters of asteroids in the orbital element space of semi-major axis, eccentricity, and inclination \citep[$a_p,e_p,i_p$;][]{zappala1990asteroid,nesvorny2015identification}. The resulting current catalogs of asteroid family membership contain more than 140,000 asteroids belonging to 119 separate families \citep{nesvorny2015identification}. 

\begin{figure*}[t!]
    \centering
    \includegraphics[width=\textwidth]{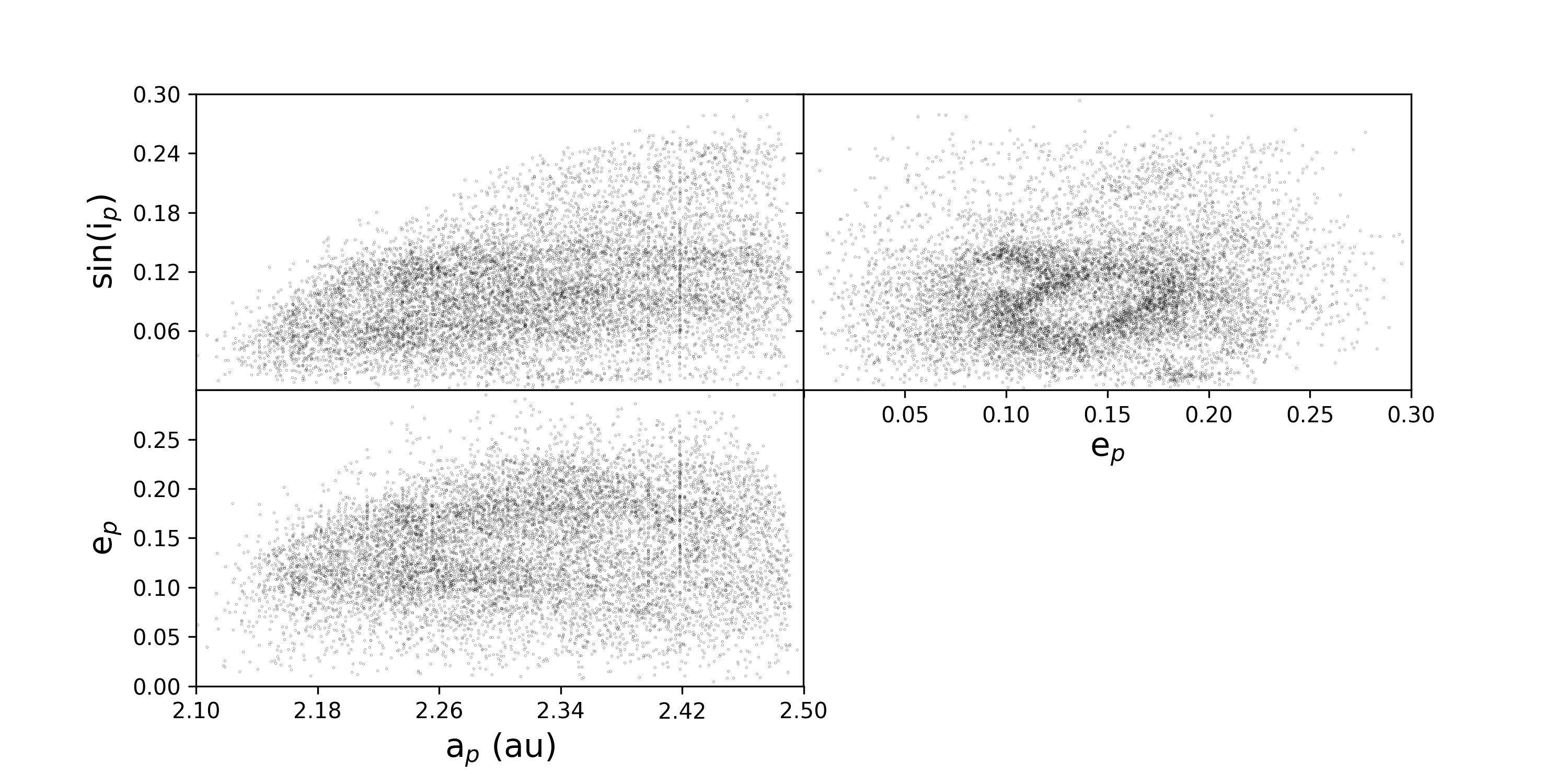}
    \caption{Same as Fig.~\ref{fig:backgroundHighAlbedo} but after removing the families identified by \citet{nesvorny2015identification}. The halos are best seen in the $\sin(i_p)$ versus $e_p$ space.  }
    \label{fig:backgroundNesvorny}
\end{figure*}

However, there are several reasons why current asteroid family catalogs are not suitable for the aforementioned cleaning of the main belt. Firstly, the HCM asteroid families are in general conservative. To demonstrate this, Fig.~\ref{fig:backgroundNesvorny} shows the so-called background asteroid population after removing all family members identified by \citet{nesvorny2015identification}. Immediately, some structures related to families are still visible, with the most obvious being the ``halos'' in the $e_p, \sin(i_p)$ plane. Family halos have also been identified by other works around several families \citep{Parker2008Icar..198..138P,broz_eos_2013}. As asteroids within the halos tend to have the same colors as the family that they surround \citep{Parker2008Icar..198..138P}, it appears logical to infer that they are constituted by asteroid fragments from the family parent body. The fact that they had not been linked to the core of the family by the HCM is a result of the fact that HCM-driven asteroid linking is kept conservative.  This conservatism is maintained in order to have good separation between adjacent families and to limit family contamination from the so-called interlopers (false positives). This is desirable when family membership information is to be used by authors interested in modeling the family dynamics or studying their composition. However, by keeping the HCM clustering conservative, many family members are missed \citep[e.g.,][]{Parker2008Icar..198..138P,broz_eos_2013}. 
The second reason is that the HCM family catalogs are likely missing families with ages roughly older than 2 Gyr \citep{brovz2013constraining,spoto2015asteroid,Bolin2017}. This deficit can be seen by comparing the number of identified families as a function of their age. Although a roughly constant asteroid break-up rate is expected over the last $\sim$3 Gyr, numerical models of Solar System evolution from \citet{Bottke2005Icar..175..111B} suggest an even higher original collision rate that decreased in time because of dynamical depletion. Such results raise doubt as to the fidelity of the low number of detected ancient families. One may hypothesize that this deficit of known old families is due to the efficiency of the HCM, which decreases with increasing age of the collisional family to identify. 

This latter effect is a result of the fact that families disperse over time. A non-gravitational effect---the Yarkovsky effect----slowly changes the orbital semi-major axis $a_p$  of asteroids at a rate $\textrm{d}a_p/\textrm{d}t$ proportional to $1/D$ \citep{vokrouhlicky2015yarkovsky}. Asteroids in prograde rotation  have $\textrm{d}a_p/\textrm{d}t>0$ and migrate towards larger heliocentric distances, whereas those in retrograde rotation with $\textrm{d}a_p/\textrm{d}t<0$ migrate towards the Sun. This creates correlations of points in the $(a_p, 1/D)$  plane called V-shapes (as they resemble the letter ``V''), whose slope ($K$) indicates family age \citep{Vokrouhlicky2006, spoto2015asteroid}. From modeled or measured $\textrm{d}a_p/\textrm{d}t$-values, family age can be determined with some uncertainty \citep{Vokrouhlicky2006, spoto2015asteroid}. Asteroids, as they move, encounter orbital resonances with the planets, which change their orbital $e_p$ and $i_p$, but not their $a_p$. Thus, families become harder to identify as clusters of points in $(a_p,e_p,i_p)$ space as they age because they are increasingly dispersed \citep{delbo2015asteroid} and overlap with each other.

However, \citet{Bolin2017} and \cite{Delbo2017} developed a method---based on the initial work of \citet{Walsh2013}---that makes use of the V-shape to identify  families. \citet{Deienno2021} studied the efficiency of this method as a function of the age of the family, its position in the main belt, and its contrast with respect to the local background. These latter authors concluded that the V-shape family identification method is particularly useful for locating strongly dispersed  collisional families $(e_p,i_p)$, that is, those not trivially visible to the HCM. This is because while $(e_p,i_p)$ of family members are dispersed by the effect of secular and mean motion resonances, these resonances do not affect the $a_p$ and therefore the correlation between $a_p$ and $1/D$ should stay visible for billions of years. 

The efficacy of the V-shape method has been demonstrated with the identification of a family that could be as old as the Solar System \citep{Delbo2017} and another that is roughly 3 Gyr old \citep{Delbo2019}. A third family, but with lower statistical significance, has also been found to be as old as our Solar System \citep{Delbo2019}. These families are quite vast in general, leading the authors to infer that most of the asteroids previously assigned to the background are instead family members.  By studying the orbital properties of asteroids in the inner portion of the main belt, \cite{Dermott2018NatAs...2..549D} came to the same conclusion; namely that most of the ``background'' asteroids are instead family members, either belonging to halos of already known families or those yet-to-be discovered. 

Previous works identified families using the V-shape methods in the inner main belt for low- and medium-albedo bodies \citep{Delbo2017,Delbo2019}. Here, we extend these works, focusing on those asteroids with intermediate and high albedo, that is, $p_V>0.12$, of the inner portion of the main belt. This population of asteroids contains several $<$2 Gyr-old families already identified by means of the HCM \citep[e.g.,][]{nesvorny2015identification}. Hence, a blind V-shape search of this population would mostly result in the detection of these known families. Therefore, here we proceed with a first additional step compared to previous V-shape searches for unknown families; namely, we reassess which bodies are members of collisional families that are already cataloged \citep[e.g., by][]{nesvorny2015identification}, and we remove them to study the remaining population. We need to perform this family membership reassessment in order to link halos to the cores of the respective families. Subsequently, we search amongst the asteroids of the remaining population for the V-shapes that represent the oldest and undiscovered asteroid families.  

Once a V-shape is identified, as following previous works \citep{Delbo2017,Delbo2019}, we then consider all asteroids outside of these V-shapes to be unassociated to any asteroid family and catalog them as original planetesimals. This is because a fragment asteroid generated by a collision in the main belt is expected to be inside a V-shape. Therefore, asteroids outside of all V-shapes cannot be associated with collisional families and are most likely pristine primordial objects. Additionally, as noted in \citet{Delbo2017,Delbo2019}, most of the identified planetesimals are large and at the same time do not reside near other large objects in (a,1/D) space, meaning that there are no potential larger objects from which the candidate bodies may originate. This affirms that these objects are unlikely to be fragments of undiscovered families that remain below a detection limit. Regardless, if some of these objects are not pristine, the collisions that produced them must have happened before the Solar System reached its current configuration and some transport mechanism that brought them to the main belt must have erased their dynamical linkage to other sibling fragments. However, it is difficult to speculate how many bodies may be fragments from ancient collisions, because the collisional rate of small bodies in this epoch of the Solar System is not well constrained. For instance, the Bern III model from \citet{emsenhuber2021new} includes the late-stage planetary formation phase and has a mass resolution of 1\% of the mass of Earth, which is still 100 times larger than the most massive asteroid, Ceres \citep{emsenhuber2021new}.

The paper is structured as follows: in section 2, we describe the input datasets; in section 3 we present the methods used for reassessing the known asteroid families and identifying undiscovered ancient asteroid families. In section 4, we present the results from applying these methods to the data. In section 5, we discuss the implications of our results on planet-formation models and the limitations of our methods of planetesimal identification, and propose future studies.

\section{Data}

All data required for this work are described below, including: the proper orbital elements, physical information, previous family definitions, and simulated data. To begin, proper orbital elements approximate quasi-integrals of the full N-body equations of motion of asteroids; they are quasi-constant in time and are the basis for family identification. Proper elements can be used for asteroid clustering \citep[see, e.g., ][for reviews]{morbidelli2002modern,Knezevic2017SerAJ.195....1K} and also for the V-shape identification \citep[see, e.g.,][]{Bolin2017}. The physical information includes $p_{V}$, diameter, and bulk density, or $p_{V}$, $D$, and $\rho$, respectively. As $p_{V}$ is indicative of physical composition, it is used to filter the dataset by helping to eliminate false positives in the family membership assignment. As we show, the mean bulk density of the asteroids comprising a given family is needed to compute the age of the family. The density contributes to the magnitude of acceleration of the Yarkovsky effect \citep{Vokrouhlicky2006,spoto2015asteroid,Bolin2017}, which depends on both the mass and volume \citep{Bottke2006}. Next, instead of searching for families with no prior information, we use their definitions from \citet{nesvorny2015identification} to aid our search. Lastly, we use a synthetic dataset of the asteroid belt background population to establish a quasi-random noise, which defines the limits of a family while using the HCM (clustering method). In the following subsections, we review the data sources.

\subsection{Orbital elements and physical properties}
We retrieved the proper orbital elements and physical properties of  asteroids from the Minor Planet Physical Properties Catalogue, MP$^3$C (mp3c.oca.eu), which is run and maintained by the Observatoire de la C\^{o}te d'Azur \citep{delbo2022EPSC...16..323D}. This database is a compilation of multiple sources. For instance, most diameters are from NEOWISE \citep{mainzer2019neowise}, WISE \citep{masiero2011main}, AKARI \citep{usui2011asteroid}, IRAS \citep{tedesco2002supplemental,ryan2010rectified}, and MSX \citep{ryan2010rectified}. Additionally, some diameters have been measured from stellar occultation campaigns \citep{Durech2011Icar..214..652D} and spatially resolved images using high-spatial resolution adaptive optics at large telescopes \citep{hanuvs2017volumes}.  The proper elements included in MP$^3$C are taken from the Asteroids Dynamic Site \citep{knezevic2012asteroids}. 

\begin{figure}
    \includegraphics[width=\linewidth]{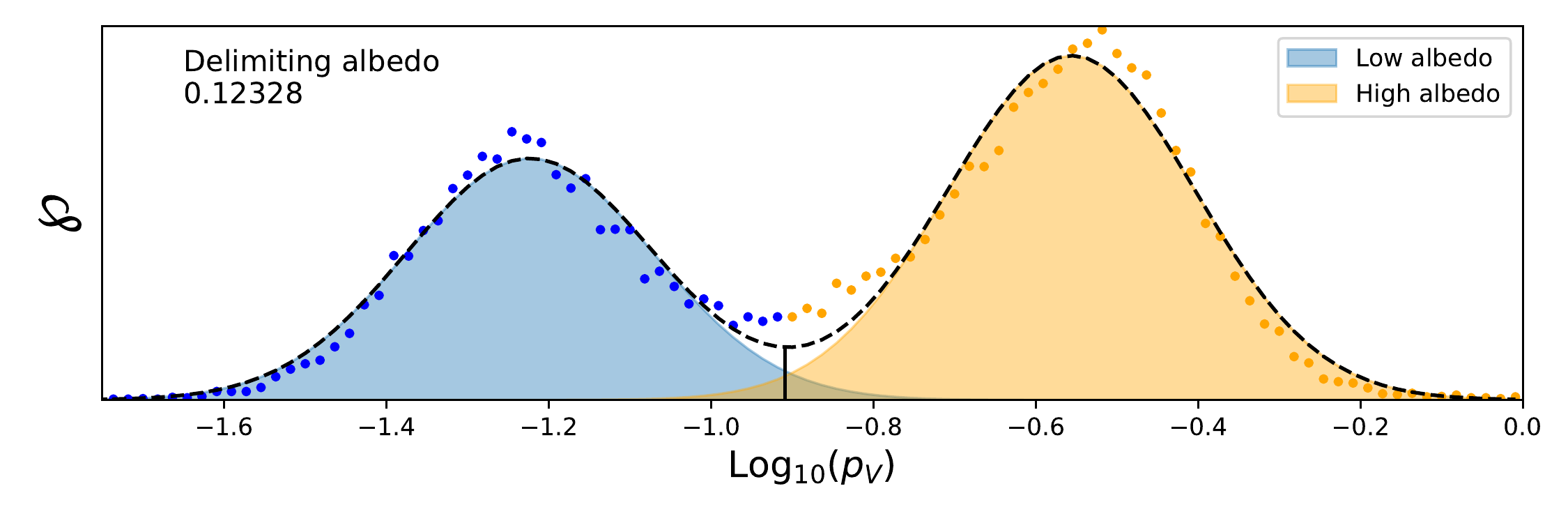}
    \caption[]{ The $\log_{10}$ distribution of $p_{V}$ for all bodies in the inner main belt. The data points are the bin heights of the histogram, the dotted black line is the best fit of the double-peaked Gaussian histogram. The unimodal components of the low and high $p_{V}$ are shown in blue and orange, respectively. We impart a delimiting $p_{V}$, which is indicated with the vertical black bar. }
    \label{fig:albedo-divider}
\end{figure}

Next, when viewing the distribution of log$_{10}(p_{V})$ of main belt asteroids, we note that there are two peaks, which are indicative of two major compositional classes. To first order, these peaks approximate the C-types and S-types. In this work, we are interested in studying the S-type population and eliminate all C-types of the inner main belt. To do this, we first distribute log$_{10}$($p_V$) into 165 bins, which is the square root of the number of data points. We then use a $\chi^2$ least squares minimization to fit two Gaussian peaks to the log$_{10}$($p_V$) distribution as:
\begin{equation}
\mathcal{P}(x) = \mathcal{P}_0(x;A_0,x_0,\sigma_0) + \mathcal{P}_1(x;A_1,x_1,\sigma_1),
\end{equation}
where $\mathcal{P}$ is the probability density, $\mathcal{P}_0$ is the probability density function of the low-$p_{V}$ population and $\mathcal{P}_1$ is of the high-$p_{V}$ population, $x$ is the logarithm of $p_V$, $A$ is the amplitude, $x_0$ and $x_1$ are the means, and $\sigma$ is the standard deviation. The histogram and fit are shown in Fig.~\ref{fig:albedo-divider}. From these, we find the delimiting value of the $p_{V}$ by finding the ratio of $\mathcal{P}_0$/$\mathcal{P}_1 =1$, which is when $p_{V}$ is 12.328\%. This value is consistent with previous findings \citep{Delbo2017}.

We consider this to be a fair delimiting $p_{V}$ because we have a small degree of contamination and loss. For instance, 1.20\% of the area of the low-$p_{V}$ peak lies above the delimiting $p_{V}$ and 1.28\% of the high-$p_{V}$ peak lies below it. In this work, we select asteroids whose $p_{V}$ is higher than 12.328\%.

\begin{figure*}[t!]
    \centering
    \includegraphics[width=\textwidth]{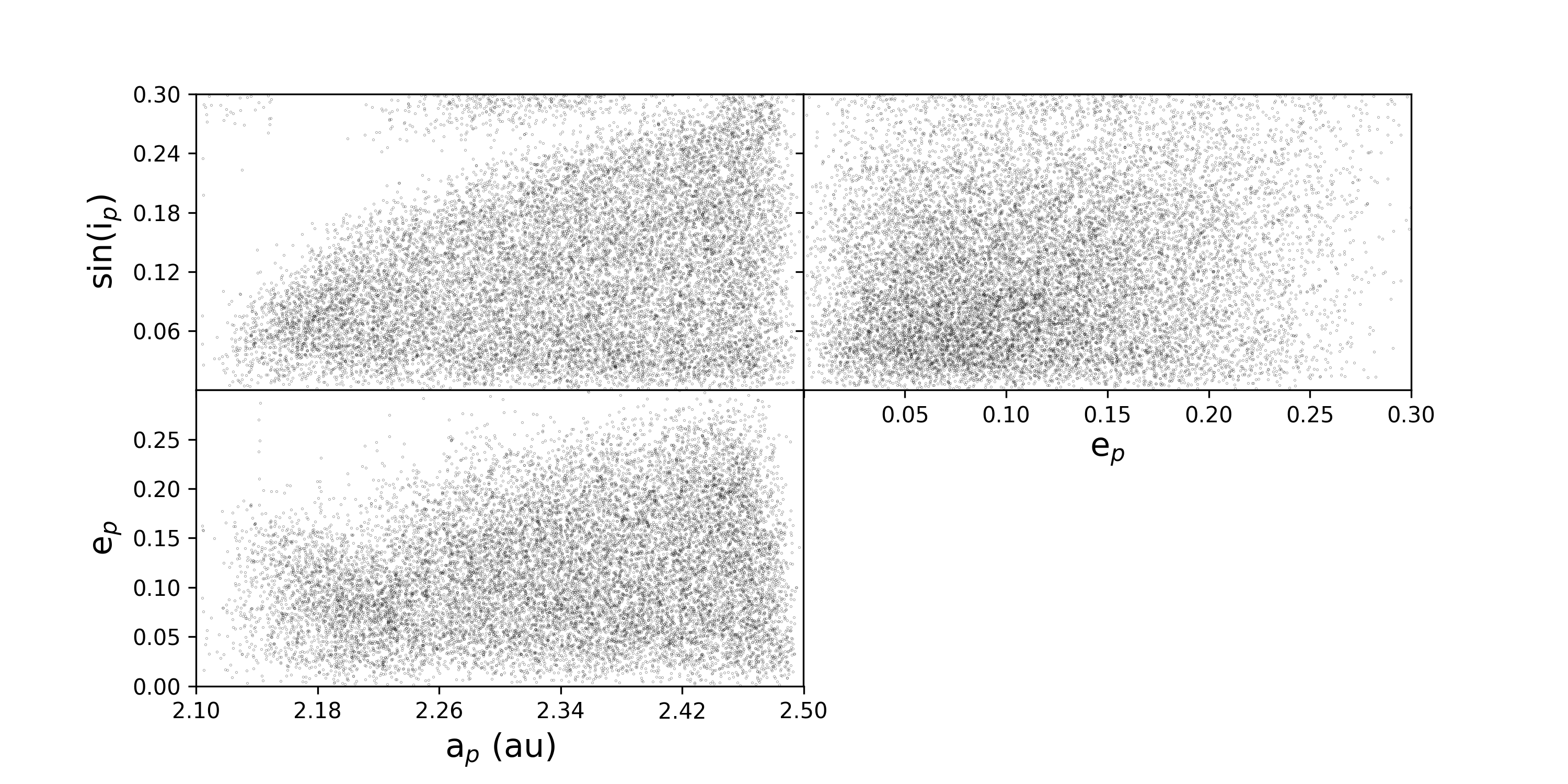}
    \caption{Axes are the same as in Fig.~\ref{fig:backgroundHighAlbedo}. The data come from the modeled background of the inner main belt from \citet{Deienno2021}, which simulates the dynamic evolution of the asteroids of the background population in the main belt. }
    \label{fig:backgroundDeienno}
\end{figure*}

\subsection{Synthetic main belt background}

We use the synthetic dataset of asteroid diameters and proper orbital elements generated by \citet{Deienno2021}, which predicts what the inner main belt would look like if no asteroid collisional families were present. To do this, \citet{Deienno2021} built on the work of \citet{tsirvoulis2018reconstructing}, who removed all of the families in the ``pristine zone'' of the main belt between 2.82 and 2.96 au, which contains easily identifiable families. After removing these families, \citet{tsirvoulis2018reconstructing} found that the remaining background population has a cumulative size--frequency distribution (SFD) described by $c(D) \propto D^{-q} $, where $q$ is the power-law slope, $c$ is the cumulative counts, and $D$ is the diameter. The value of $q$ was found to be 1.43 and \citet{Deienno2021} assumes this to be the same power law governing the background population of the entire main belt. While generating bodies from this distribution, the authors truncate the diameters at a minimum size of 2 km. This is still valid for our purposes because 70\% of the bodies of the inner main belt from the MP$^3$C have diameters greater than 2 km and none have diameters of smaller than 1 km. Additionally, the slope of the V does not change with diameter, nor is diameter an input to the HCM.

After their generation, the synthetic orbital elements of \citet{Deienno2021} are randomized and then integrated for 100~Myr. With this, the bodies felt gravitational effects of all solar system planets from Venus to Neptune and the non-gravitational Yarkovsky thermal force. The authors observed that the steady state was reached by 100~Myr and continuing the evolution only depleted the population. Lastly, instead of performing a detailed Hamiltionian analysis to obtain the proper orbital elements, the authors opted to average the Keplerian orbital elements over the last 10~Myr as a proxy \footnote{For asteroids with proper eccentricity or inclination larger than the forced eccentricity or inclination, respectively, the time-averaged eccentricity or inclination is close to the proper value. The difference is instead large if the proper eccentricity or inclination is smaller than the respective forced one. Because the forced eccentricities and inclinations are small, the differences between averaged orbital elements and the proper ones is small for most asteroids.}. Although different in definition, the simplified time averaging is more than sufficient for our purposes since we are not interested in the precise orbital history of individual simulated asteroid orbits, but the ensemble statistical properties of the main belt. The final result of \citet{Deienno2021} is displayed in Fig.~\ref{fig:backgroundDeienno}.

\subsection{Usage of the known families}
We use the definitions of the high-$p_{V}$ inner main belt families from \citet{nesvorny2015identification} as a starting point. These families are Baptistina, Flora, Lucienne, Massalia, Nysa, and Vesta. This is a simpler picture of the families in this region because we are not considering the complex subfamily structures. For instance, there is evidence of a subfamily within Nysa whose largest object is (135) Hertha \citep{Dykhuis2015Icar..252..199D}. Given that we are searching to remove all known families, this problem does not concern us, because removing Nysa will also remove Hertha. 

The parameters describing the V-shape are taken from \citet{nesvorny2015identification}. More precisely, we take the $C_0$ -values, which can be related to the slope, $K$, of the V-shape by $K=\sqrt{p_V}/(1329 C_0)$. Additionally, when we apply the HCM, we search for the families one by one, instead of searching for multiple clusters by applying the HCM to the entire dataset, as has been done in previous works \citep{masiero2013asteroid}. While searching for one cluster at a time, the HCM must be given a central body from which to start its search. The central bodies of the families are also taken from \citet{nesvorny2015identification} and are typically the largest asteroid in a given family. Given this, we proceed to our own family membership assessment in order to be more inclusive, following an approach similar to that of \citet{tsirvoulis2018reconstructing}, but with important differences which are explained in the methods section that follows.

\section{Methods}\label{sec:methods}
Given the problem that the current asteroid family membership lists---found with HCM---are conservative, as detailed in Sect. \ref{S:intro}, our first procedure is to reassess the extent of known families using a new method that attempts to merge their cores with their halos. This consists in constraining the HCM using the V-shapes of the families and a statistically robust method to stop HCM clustering. Once we have removed the reassessed known families, we hunt for missing ones---which are likely older than $\sim$2.5 Gyr---by searching for their Yarkovsky effect signature \citep[i.e., their V-shape;][]{Bolin2017}. As we demonstrate below, this search is successful.

This experiment is procedural, which is to say that results from one step are the initial conditions for the next. In order to estimate uncertainties, at each step, we explore all possible values for all free parameters with a Monte Carlo approach and restrict the  range of each free parameter to a valid domain. This allows us to know the sensitivity of our results with respect to the free parameters. Here, we layout the assumptions and actions taken in each step to quantify and limit our uncertainty. First, the underlying assumptions are as follows, to first order:
\begin{enumerate}
    \item The asteroid belt is composed of collisional families, which are amongst the planetesimals that comprise the ``background'' population.
    \item Members of a collisional family cannot drift farther in the semi-major axis than allowed by the maximum Yarkovsky drift rate; if asteroids encounter zones of dynamical instability  during this process, then these bodies are likely removed from the main belt.
    \item The SFD of planetesimals in the pristine zone of the main belt \citep[as estimated by][]{tsirvoulis2018reconstructing} is the same as the SFD of the inner main belt.
    \item The $p_{V}$ of an asteroid is indicative of composition, and asteroids of different composition do not originate from the same parent body. 
\end{enumerate}

A significant consequence of these assumptions is that all bodies outside all V-shapes are planetesimals. The last three assumptions clearly have limitations, but we consider them valid for the scope of this work. For instance,  collisional families containing members of distinct composition have been reported \citep{oszkiewicz2015differentiation,Fornasier2016Icar..269....1F}; but in general, members have uniform composition within each family \citep{nesvorny2015identification}. With these assumptions in mind, we can outline our methods, showcasing where and how we treat the uncertainties and free parameters. To specify, we define uncertainties as quantities that have associated errors bars, namely the diameter, $D$, and its 1$\sigma$ uncertainty, $\delta D$. The free parameters are subjective parameters that take a fixed value per calculation, affect the result, and whose best value cannot be known beforehand. The two free parameters considered here are the absolute number density of the synthetic background $\text{N}_{d}$, and the width of the lobes used in the V-shape detection, $a_w$ \citep[see][Fig.~3]{Bolin2017}. Finally, when searching for the original SFD, we consider two different models to correct for the depletion of bodies as a function of diameter via dynamical and collisional loss.

\subsection{The V-shape-constrained hierarchical clustering method}\label{sec:V-shape-constrained-hcm}

Here, we present a new technique for rejecting more false positives than was possible before and for determining the {quasi-random noise}, which is an inherent problem when using the HCM \citep{zappala1990asteroid}. We briefly describe the input and output datasets and the variables involved in this step of the experiment. We then describe our method in detail, while in parallel we describe how clustering was done in previous works. 

\begin{enumerate}
    \item Inputs:
    \begin{itemize}
        \item $\{a_{p},e_{p},i_{p},D\}_{\text{Real}}$ . These are proper elements and diameters for all high-$p_{V}$ asteroids in the inner main belt within the MP$^3$C catalog.
        \item $\{a_{p},e_{p},i_{p},D\}_{\text{Syn}}$ . These are proper elements and diameters of the synthetic dataset from \cite{Deienno2021}.
    \end{itemize}
    \item Free parameters considered here:
    \begin{itemize}
        \item $[\text{N}_{d}]$ , which is the absolute value of the number density of the synthetic background.
    \end{itemize}
    \item Uncertainties considered here: 
    \begin{itemize}
        \item $\{\delta a_{p},\delta e_{p},\delta i_{p}\}_{\text{Syn}}$. These are the proper elements of the synthetic background while applying the clustering algorithm.
    \end{itemize}
    \item Outputs:
    \begin{itemize}
        \item $\{a_{p},e_{p},i_{p},D\}_{\text{FR}}$ . This is the ``Family Removed'' dataset, which is a subset of $\{a_{p},e_{p},i_{p},D\}_{\text{Real}}$.
    \end{itemize}
\end{enumerate}

To build from the start, the classical HCM works by first placing each body in its own cluster and then subsequently merging clusters as a given distance parameter is increased \citep{zappala1990asteroid}. In the case of asteroid families, this distance metric is a velocity, $v$, which can be understood as the $\Delta v$ required to change the  orbit of one asteroid to that of another. The velocity that is used to define the size of the family is known as the velocity cut off \citep[$v_c$,][]{zappala1990asteroid,nesvorny2015identification}. Deciding the value of $v_c$  has known difficulties. For instance, at small $v_c$, a family cluster clearly contains too few members. On the other hand, as $v_c$ is increased, the HCM will include more family members but at the same time will include bodies that belong to other families or the background. Indeed, if $v_c$ is large enough, the entire dataset would be grouped into one family, mostly including false positives into the real family. 

The ideal situation would be to increase the $v_c$ as much as possible without accepting any false positives. One excellent way to do this is to reject all bodies that cannot belong to the family {a priori}, and restrict the HCM to search for a cluster over a smaller subset of data that only includes plausible family members. 

The first criterion used in this study to reject false positives is the $p_{V}$. To reject false positives even further, we introduce a new technique, which is to constrain the HCM to cluster asteroids that are only inside the V-shape of a particular family. That is to say, we subsample the real dataset, $\{a_{p},e_{p},i_{p},D\}_{\text{Real}}$, based on the semi-major axis and the diameter. Qualitatively, all bodies outside of the target family’s V-shape are rejected. Quantitatively, the V-shape of a family is determined by the bodies that experience maximum Yarkovsly drift, the diameters of such bodies given with the relation: $1/D = K|a_{p} - a_c|$, where $a_c$ and $K$ are the vertex and slope parameters of a family’s V-shape, and $a_p$ and $D$ are of an asteroid undergoing maximum drift rate. Then, asteroids (denoted with index j) are rejected if $1/D_j < K|a_{p,i} - a_c|$ and accepted if $1/D_j > K|a_{p,i} - a_c|$. For each family, we do not redetermine $a_c$ and $K$ but simply take them from the compiled list from \citet{nesvorny2015identification} and use them as fixed values. We find that this is a viable method because the V-shape parameters are in principle independent of the number of family members added \citep{Milani2014Icar..239...46M} meaning that, the V-shape should not change by reassessing family membership. 

Most bodies that are outside of the V-shape should be unrelated to the family, because the Yarkovsky drift is not able to carry them that far; we say most, because some phenomena, such as impulse-like events including close encounters with massive asteroids (and planets in rare cases), can push asteroids outside of the Yarkovsky V-shape---not to mention that measurement uncertainties may also cause a small percentage of members to be counted as being outside of the V-shape. These asteroids that experienced impulse-like events had their orbits dramatically changed over short timescales, which broke their dynamical link to the family. This makes it impossible to differentiate them from background asteroids and relate them to a specific family on the basis of dynamics alone. Fortunately, it has been shown that the effect of these processes is small \citep{Delbo2017}.

We note that, within a V-shape, all values of $e_p$ and $i_p$ are accepted; therefore, while viewing the subset of data in orbital element space for a particular family, $a_{p}$ will be restricted but $e_{p}$ and $i_{p}$ will span their entire respective domains. With this subset, we apply the HCM to the central body of each family and study the growth of the number of its members as a function of $v_c$.

The second novelty of the present work is the method we use to determine the most optimal $v_c$. We extend the idea of a ``quasi-random noise'', which was introduced by \citet{zappala1990asteroid}. We apply the HCM at sequentially increasing values of $v_c$ to both a real and a reference dataset in parallel. The reference dataset is a random distribution of points that should have no clusters. When the HCM begins forming clusters on the reference dataset, the $v_c$ has been determined. This is because the algorithm is clustering objects from statistical chance, and not because a cluster exists there. 

\begin{figure*}
    \includegraphics[width=.35\linewidth]{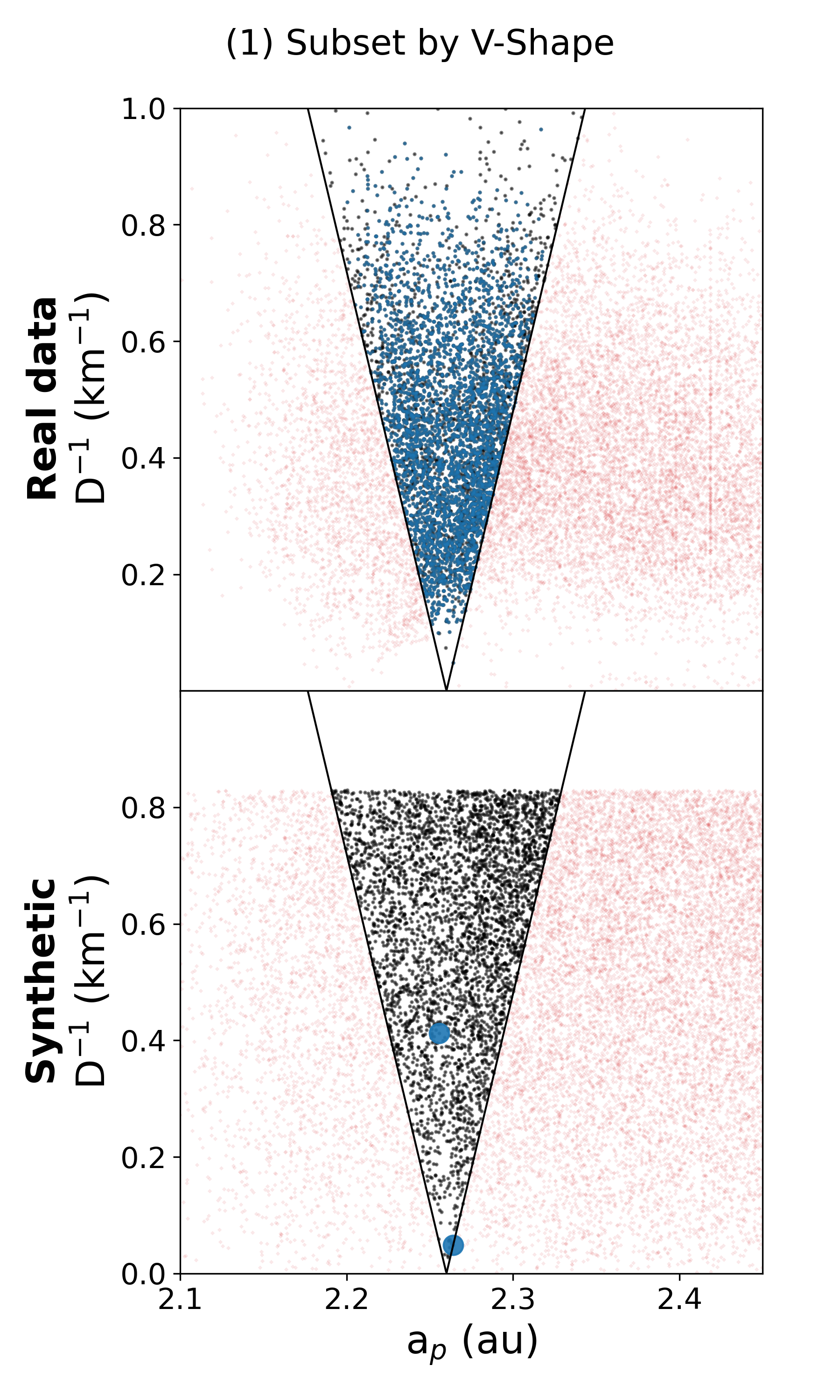}
    \includegraphics[width=.6\linewidth]{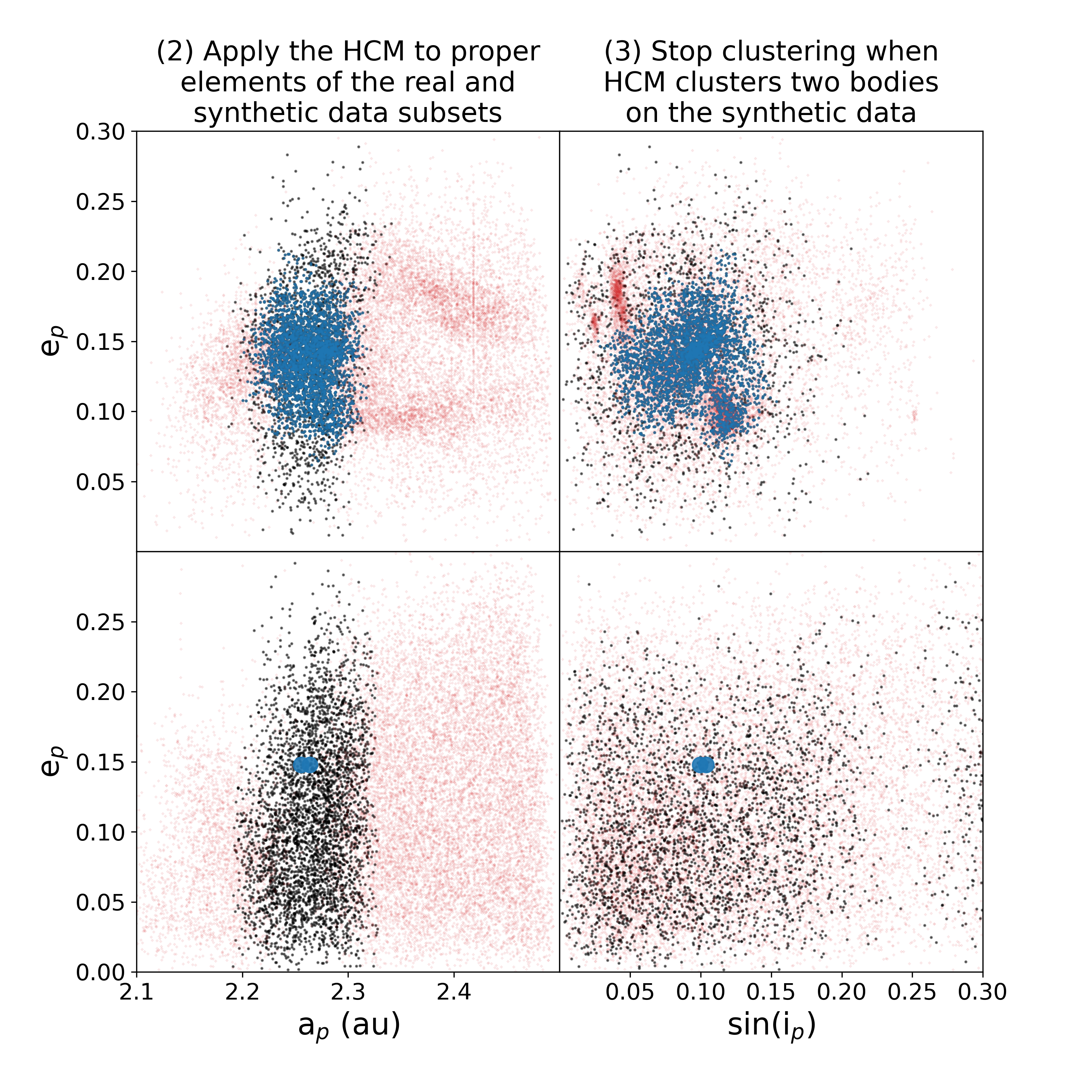}
    \caption{Demonstration of the V-shape-constrained HCM. The top row depicts the real data while the bottom shows the synthetic background of \citet{Deienno2021}. The red bodies are a priori rejected since they are exterior to the family’s V-shape. The blue bodies are those clustered by the HCM. In the bottom panels, the two bodies are enlarged for clarity and only two bodies are clustered because forming a cluster on the synthetic dataset determines the $v_c$ (clustering threshold) for the real data. Thus, the blue bodies in the top plot correspond to family members. The black bodies are those who were not a priori rejected but also not clustered at the given $v_c$. The plots on the right show the same dataset as the left column but plotted in proper element space, which is the data space to which the HCM is applied.}
    \label{fig:baptistina-vshape}
\end{figure*}

Previously, \citet{zappala1990asteroid} generated the reference dataset by shuffling the real data in such a way as to dissolve the clusters while preserving large-scale variations in number density. Though ground-breaking  at the time, we are now able to improve on this by taking advantage of a synthetic dataset generated by \citet{Deienno2021}. To determine the quasi-random noise, we subsample the synthetic main belt using the V-shape of each family  in the same way as the real dataset, and then add the central body. The HCM then searches for a cluster about the central body. The $v_c$ is determined when two bodies are clustered in the synthetic data set. This entire process is demonstrated in Fig.~\ref{fig:baptistina-vshape}.

By introducing the synthetic dataset, in addition to inserting a modeling assumption, as listed in Sect.~\ref{sec:methods}, we introduce a free-parameter that will impact the results. That is, although the spatial variation of the synthetic dataset is well modeled, the absolute value of the count density, $\text{N}_{d}$, is arbitrary. 
We therefore perform a sensitivity analysis on $\text{N}_{d}$, spanning a range from physically too large to excessively small. First, we start with 18,191 bodies, which is the number of bodies provided from \citet{Deienno2021} and conveniently greater than the size of the real dataset, which has 16,200 bodies. We choose a lower limit of $\text{N}_{d}$/100, which only contains 182 bodies. This intuitively appears to be too few, which we later show to be the case. Within this range, we explore several densities as a fraction of the original number of bodies, as listed in Table~\ref{tab:hcm-simulated-bg}.


\begin{table*}
    \caption{The $v_c$ at which the HCM clustering is stopped for each family for different N$_{d}$ of the artificial background in units of m~s$^{-1}$.}\label{tab:hcm-simulated-bg}
    \begin{center}
    \begin{tabular}{|l|rrrrrrrrr|}
    \hline
    N$_{d}$ (frac.)  & 1   & 3/4 & 2/3 & 1/2 & 1/3 & 1/4 & 1/5 & 1/10 & 1/100 \\
    \hline
    Baptistina & 150   & 150          & 150       & 150  & 165   & 165    & 175   & 200   & 540      \\
    Euterpe    & 80    & 80           & 80        & 80  & 115    & 225    & 160   & 280   & 540      \\
    Flora      & 200   & 200          & 200       & 200  & 225   & 225    & 225   & 265   & 547      \\
    Lucienne   & 95    & 95          & 95        & 95  & 110    & 180    & 180    & 252   & 667      \\
    Massalia   & 135   & 135          & 135       & 135  & 140   & 185    & 195   & 215   & 497      \\
    Nysa       & 140   & 140          & 140       & 150  & 150   & 190    & 195   & 215   & 450      \\
    Vesta      & 145   & 145          & 145       & 145  & 180   & 205    & 205   & 245   & 517  \\   \hline
    \end{tabular}
    \end{center}
    \tablefoot{
        These are the median values found from applying the HCM to 100 realizations of the artificial background as shown in Fig.~\ref{fig:velocityCutOffHistogram}. The columns correspond to uniformly decreasing the N$_{d}$ of the artificial background to fractions of the original total number of bodies, 18,191, as explained in section~\ref{sec:V-shape-constrained-hcm}.
    }
\end{table*}

Next, to create each subset, we randomly removed bodies via their index until we lowered the original number density to the desired fraction, ensuring that we do not alter the relative density variations of the proper elements. Yet, this random subsampling adds a stochastic uncertainty as we apply the HCM. That is to say, in one random subsampling, an asteroid could be placed in close proximity to the central body, providing a smaller $v_c$, yet in another case the closer bodies could be removed, leading to larger values of $v_c$. To combat this, for each density we perform 100 random subsamplings and apply the HCM to each of those. Then, we examine the distributions of $v_c$ for each $\text{N}_{d}$, which is presented in Fig.~\ref{fig:velocityCutOffHistogram} of the Appendix. One thing to notice is that for all densities greater than 1/10, the distributions are well behaved, whereas when the background number density is 1/100 there is a seemingly uniformly random selection of $v_c$, indicating that this $\text{N}_d$ is too low. Given this, we then only consider densities between 1 and 1/10 for the remainder of the study. At each selected density, we choose the median value of the distributions of the $v_c$-values as the final velocity cut-off for each family. 

Next, we apply the HCM---constrained by the V-shape on the real data using the $v_c$ obtained at the previous step---to each family of Table~\ref{tab:hcm-simulated-bg}. After performing the reassessment for each family, we removed all objects identified as collisional fragments. As an output from this step, we have eight family-removed datasets: $\{a_{p},e_{p},i_{p},D\}_{\text{FR}}$, one for each $\text{N}_d$ considered. By doing so, for the next steps where we search for undetected families, we can also explore the sensitivity of our final results to the free parameter $\text{N}_d$.

\subsection{V-shape search for ancient families}\label{sec:vsearch}
After reassessing and removing the known families, we are left with a ``background'' in which there are no obvious signs of clusters of families in proper element space (as can be seen in Fig.~\ref{fig:proper-elements-no-families}). In an effort to discover previously obscured families, we search for traces of them using the V-shape detection method \citep{Bolin2017,Delbo2017}. By continuing from the family-removed background from the previous step, we apply a modified V-shape detection scheme that considers the following:
\begin{enumerate}
    \item Inputs:
    \begin{itemize}
        \item $\{a_{p},e_{p},i_{p},D\}_{\text{FR}}$ , which is the family removed background.
    \end{itemize}
    \item Free-parameters:
    \begin{itemize}
        \item $[\text{N}_{d}]$, the absolute number density of the synthetic background.
        \item $[a_{w}]$, the width of the lobes in the V-shape search method.
    \end{itemize}
    \item Uncertainties: 
    \begin{itemize}
        \item $\{\delta D\}_{\text{FR}}$, the uncertainties in the asteroid diameters.
    \end{itemize}
    \item Outputs:
    \begin{itemize}
        \item $\left(a_{c}\pm\delta a_{c},K\pm\delta K\right)$, V-shape parameters of the detected  families.
        \item $t_{\text{age}}\pm\delta t_{\text{age}}$,  the age of the detected families.
        \item X-$\sigma$, the probability that the ancient family detection is not a statistical coincidence.
        \item $\{a_{p},e_{p},i_{p},D\}_{\text{AF}}$,  members of the V-shape detected families.
        \item $\{a_{p},e_{p},i_{p},D\}_{\text{P}}$, planetesimals.
    \end{itemize}
\end{enumerate}    

In summary, the V-shape detection method works by detecting a change in density in the $(a,1/D)$ plane while transitioning from a region occupied by the family to a region unoccupied by the family. In other words, the entire space is scanned with a series of test vertices and slopes of the V-shape, which creates a two-dimensional grid known as the score-map. There are a few different ways of performing this scan, though we use the $a_w$ method, which was developed by \cite{Bolin2017} and \cite{Delbo2017} and further studied by \cite{Deienno2021}.

In this method, two lobes are created interior and exterior to the best-fit line \citep[see Fig.~3 of][or  Fig.~\ref{fig:ancient-family-detection} in this work for a visualization of the lobes]{Bolin2017}. These two lobes are bounded by the best-fit line and other V-shape lines that have the same slope and that are merely shifted above or below by user-defined distance in semi-major axis, $a_w$. The interior lobe is therefore between the line y=K$(|a-a_c| + a_w)$ and the best-fit line, whereas the exterior lobe is between y=K$(|a-a_c| - a_w)$ and the best-fit line. The $y$ for the best-fit line (y=K$|a-a_c|$) has units of 1/$D$ and would be an asteroid that is experiencing maximum Yarkovsky drift.

To measure the change in number density, we count the number of bodies within the interior and exterior lobes. (We note that this is a number density in (a,1/$D$) space and not the number density, N$_d$, in proper element space.) The ratio found between the number of interior bodies squared, $N_{\text{int}}^2$, and the number of exterior bodies, $N_{\text{ext}}$, is used as the score. A grid of test values is created that spans a range of both $a_c$ and $K$. A score is calculated at each coordinate. Local maxima of this score map are candidate detections of family V-shapes. 

In this study, we are interested in detecting the oldest families in the inner main belt that are likely older than about two billion years \citep[which are thought to be beyond the effectiveness of the HCM; see][]{brovz2013constraining,spoto2015asteroid,Bolin2017}, and therefore we search for slopes smaller than 1.5~km$^{-1}$~au$^{-1}$. Given the age estimation from \citet{nesvorny2015identification}, we calculate the age of an asteroid family as:
\begin{equation}\label{eq:age-estimation}
    \begin{split} 
        t_{\text{age}} \approx 1 \text{Gyr} \times \left(\frac{1}{5}\right)^{1/2}\left(\frac{1}{1329 km}\right) & \left(\frac{1}{10^{-4}au}\right) \\
        \times \left(\frac{a_{c}}{2.5 au}\right)^2 & \left(\frac{\rho}{2.5 g cm^{-3}}\right) \left(\frac{1}{K}\right),
    \end{split}
\end{equation}\label{EQ:age}
where $\rho$ is the mean bulk density of all asteroids in the family. A family with a slope of 1.5~km$^{-1}$~au$^{-1}$ and a mean bulk density of 2.5 g~cm$^{-3}$ placed in the inner main belt would have an age of 1.9 Gyr. Therefore, covering all slopes smaller than 1.5$^{-1}$~au$^{-1}$ ensures that we scan for all families that are likely beyond the HCM detection limit \citep{Bolin2017}. Additionally, we use Eq.~\ref{eq:age-estimation} for computing the age of detected families and accordingly find its error as:
\begin{equation}
    \sigma_{t_{\text{age}}}^2 \approx \left( \frac{\partial t_{\text{age}}}{\partial a_{c}} \right)^2\sigma_{a_c}^{2} + \left( \frac{\partial t_{\text{age}}}{\partial K} \right)^2\sigma_{K}^{2} + \left(\frac{\partial t_{\text{age}}}{\partial \rho }\right)^2\sigma^{2}_{\rho} + 2\frac{\partial t_{\text{age}}}{\partial a_c}\frac{\partial t_{\text{age}}}{\partial K}\sigma_{a_{p}K}^2,
\end{equation}
where $\sigma$ represents the standard deviation of the corresponding variable and $\sigma_{a_c k}^2$ is the covariance between the slope and vertex.

As we show in section~\ref{S:Results}, our V-shape scan results in a detection of one of the most ancient families known. We want to gain some insight into the membership of the ancient family, which is unfortunately beyond the HCM detection limit. To bypass this, we consider bodies within the interior lobe as being family members. These clearly comprise a fraction of the entire family, that is, the few members that have near-maximum Yarkovsky drift, and there should be many more family members interior to the V-shape. Unfortunately, from the data we consider here alone, we are unable to discern them from other bodies that are in the V-shape, notably: planetesimals or missed bodies from younger clustering families, known or unknown. In fact, this proxy of using the interior lobe to probe membership is the reason we chose the $a_w$ method. Previous works also considered the $dK$ method, which, instead of defining the lobes by two other lines that have the same slope with a slight vertical offset, defines the lobes with two lines that have slightly different slopes but share the same vertex. As the lobes come to a point at the vertex, bodies with small $1/D$ may not be interior to the lobe, and may therefore be excluded from the membership list. 

To obtain a detection of such a family, we followed a three-step process, considering the free parameters: N$_{d}$ and $a_{w}$ and the uncertainties in the diameters. Again, for each of these we performed a Monte Carlo-style analysis. After these procedures, we identified a family, and we report a confidence level of its detection in section~\ref{detectionofanancientfamily} and the resultant data products in Appendix~\ref{appendix:ancient-primordial-members}, which include identification of some planetesimals below its V-shape as well as some members of the ancient family. 


\subsubsection*{Step 1: Performing a global search}

\citet{Deienno2021} investigated the robustness of a series of techniques for identifying asteroid families, one of which is the $a_w$ technique. The authors find some key results: first, families with a higher contrast to the background are clearly easier to detect. Second, the score map (see Fig.~8 of \citealt{Deienno2021} or Fig.~\ref{fig:global-search} here for a visualization) should show a curved $X$-like shape when a family is detected. Each line of the $X$ corresponds to the right or left sides of the lobes of the V-shape search, and the maximum score should be at the intersection of the two lines of this $X$. Next, and most importantly, the optimal $a_w$ anti-correlates with the slope of the family; that is, larger values of $a_w$ allow improved detection of more open V-shapes.

However, \citet{Deienno2021} were able to calibrate the $a_w$ method with the simulated data because they knew the correct values of the vertex and slope they were aiming to detect, which is not the case for us. To combat the subjectivity of the choice of $a_w$, we test a large range from too thin to too thick, and then restrict $a_w$ over a reasonable domain. More precisely, for the global search, we use nine different values of $a_w$ logarithmically spaced from 0.01~au to 0.5~au. The lower limit creates lobes that are too thin, and these therefore only capture a small number of bodies per $N_{\text{interior}}^2 / N_{\text{exterior}}$, and the search is sensitive to stochastic placement of the asteroids in $(a_p,1/D)$ space. In other words, the sampling using thin lobes is more sensitive to the random placement of bodies than the larger scale variation of where a family begins and ends. The larger limit is 0.5~au, which is the entire width of the inner main belt. As this upper limit is of the same order as the width of the domain of semi-major axis values for our entire dataset, it would not be sensitive to variations within this domain. 

To specify the technical details of our procedure, first we begin by defining the scan range, where each score map is a 100$\times$100 linearly spaced grid, where the vertex varies within 2.1 < $a_c$ < 2.5 au and the slope varies within the range of 0~<~$K$~<~1.5 km$^{-1}$~au$^{-1}$. We then begin creating base datasets in a Monte Carlo fashion by considering the uncertainties on the diameter, as well as the free parameter $\text{N}_d$. That is, for each of the seven densities ranging between 3/4 and 1/10 in Table~\ref{tab:hcm-simulated-bg}, we create 20 random realizations of the asteroid diameters based on Gaussian sampling, where the best value is the mean and the uncertainties are one standard deviation\footnote{We note that about 10\% of the diameters have fractional uncertainties of greater than 1/2, and only a handful are greater than 1. These high fractional uncertainties result in some iterations producing asteroids with negative diameters, which is a nonphysical consequence of approximating the errors as being Gaussian distributed. To avoid this, we take the absolute values. A more thorough treatment could include a bayesian inference of the diameter posterior probabilities given the infrared observations. The limitation arising from this approximation is negligible because the fraction of high-uncertainty asteroids is low, and the diameters  of those bodies are small. Small bodies are ultimately not categorized as planetesimals, which are typically large.}. In all, we create 140 base datasets. Next, we fill in the score maps of each of the 140 base datasets for all nine $a_w$-values, leading to 9x140=1,260 score maps. The 140 datasets are then averaged for each $a_w$ and we are left with nine total score maps to evaluate for our global search.

\subsubsection*{Step 2: Fine-tuning the detection}

With the nine score maps ready, we first evaluate each by eye. If a family is present, we expect that as $a_w$ increases, the $a_c$ of the local maxima remains constant and $K$ decreases. After the candidate family is identified, we then begin fine-tuning the detection by performing a second search over a restricted range of $a_c$, $K$, and $a_w$. We restrict $a_w$ because we anticipate only a small range will converge to the proper detection, as demonstrated by \citet{Deienno2021}. We also desire to use the smallest $a_w$ possible. We anticipate that larger $a_w$ will work best for detecting old families. Nevertheless, we use the widths of the inner lobe as a way to gather family members. In this way, we want the thinnest lobes possible to limit the number of false positives. 

In regards to reducing the search range of $a_c$ and $K$, the motivation is simple. We expect in the global search that a family may be indicated by a local maximum on the score map---which may not necessarily be the global maximum. Therefore, we restrict the search range of $a_c$ and $K$ such that the local maximum of the family is also the global maximum. We opt for this instead of applying another algorithm to sort all local maxima and select the correct coordinates.

Next, after $a_w$ and the search range of the score map are restricted, we perform another search against all 140 base datasets. We consider bodies within the interior lobe to be family members, and those below it to be planetesimals. To establish a confidence of our membership, we track the number of times a body is assigned to either the planetesimals or the ancient family. Bodies that are assigned more often to one class than another are more likely to be true members.  Thus, we can choose our membership as a function of statistical significance. In this work, we report bodies at a statistical significance level of 1$\sigma$. That is, bodies that are assigned for more than 68\% of the total number of iterations.

\subsubsection*{Step 3: Quantify detection confidence}\label{sec:detection-confi}

Finally, we employ a statistical test to establish detection confidence by following the same procedure as \citet{Delbo2019}. Briefly, we test the null hypothesis that the 1/$D$ distribution has no correlation with the distribution of semi-major axis. Here, as in the previous steps, we perform a Monte Carlo approach, considering many base datasets by varying N$_{d}$ and performing many realizations of the diameters. Effectively, we compare multiple base datasets---which serve as our nominal case---to multiple {``shuffled''} datasets. To create the shuffled datasets, we first find the distribution of semi-major axis, $\mathcal{D}$(a$_p$), and subsequently randomly sample this distribution for each asteroid. Between each nominal and shuffled dataset, we compare the ratio of the number of bodies that are inside of the family's V-shape, $R_{\text{in/out}}$ to those outside. In effect, if the detection is real, $R_{\text{in/out}}$ will be systematically larger in the nominal case than in the shuffled case. Whereas if the detection is a coincidence, then there will be no systematic preference between $R_{\text{in/out}}$ being greater in either the nominal or shuffled cases. To quantify the detection confidence, we perform the following procedure: 

\begin{algorithmic}
\State - \texttt{counts}=0
\For{ N$_d$ in \{N$_d$\}}
        \State - Evaluate the distribution of the semi-major axes $\mathcal{D}$($a_p$)
        \For{j in 100}
                \State - \parbox[t]{200pt}{%
                        Build nominal set $Nom=$\{($a_p,D$)\} by Monte Carlo Gaussian sampling of the diameters: \{$D| \mathcal{N}$($D,\delta D$)\}\\ }
                \State - Measure $R^{nom}_{in/out}$
                \For{k in 100}
                        \State - \parbox[t]{200pt}{%
                        Build shuffled set, $Shuff$=\{($a_p,D$)\}, with the same diameters from the nominal set but randomly sampling the semi-major axis: \{$a_p |\mathcal{D}$($a_p$)\}\\ }
                                
                        \State - Measure $R^{shuff}_{in/out}$
                        \If {$R^{shuff}_{in/out} \geq R^{nom}_{in/out}$}        
                                \State - \texttt{counts}+=1
                        \EndIf
                \EndFor
        \EndFor
\EndFor
\end{algorithmic}
At the end of this procedure, we look at \texttt{counts}, which is the number of times the shuffled dataset has a larger ratio of the number of bodies inside to outside the V-shape than the nominal case. We then divide this number by the total number of iterations and find the statistical significance.

\subsection{Correction of dynamical and collisional loss}
By reassessing the known families as well as detecting new ones, we can remove them and obtain a list of planetesimals. At this point, we wish to infer the original SFD during the early days of the Solar System from the SFD we observe today. To do this, we replicate the methods developed in \citet{Delbo2017}. Briefly, the authors correct the SFD for the dynamical and collisional loss of asteroids in the inner main belt. To start, a computer program is given a nominal observed SFD of planetesimals. Additionally, the authors appended their list with parent bodies of other known asteroid families. Beyond this, they also corrected for the fragmentation of these bodies. In our case, we append the planetesimals with bodies such as (8)~Flora, (27)~Euterpe, and (289)~Baptistina. In the case of Flora, its current day diameter is estimated to be 130~km, but by adding the mass of its collisional fragments, we can estimate it was a body of 260~km in size. We note that this estimate uses the family membership list from \citet{nesvorny2015identification}, and not the list found in this work.

Next, the corrections for the dynamical and collisional loss were treated separately. For the collisional loss, a collisional lifetime was determined by \citet{Bottke2005Icar..175..111B}, which is a characteristic time in which a fraction of 1/$e$  of the asteroids of a given diameter are destroyed \citep[see Fig.~1 of ][for a visual of the collisional lifetime as a function of diameter]{marchi2006general}. Moving forward, estimation of collisional loss is treated in a Monte Carlo fashion, where the input parameters are the total integration time, $T$, the integration step time, $\Delta t$, the collisional lifetime as a function of diameter, $t_c(D)$, and a list of asteroid diameters \{$D$\}. At each time step, the probability of an asteroid being destroyed is approximated as $\mathcal{P}$ = $\Delta t$/$t_c$, which is valid in the limit $\Delta t$ << $t_c$. In the simulations we consider, the integration step is 10~Myr, while the collisional lifetime of asteroids greater than about 35 km is roughly 4.5~Gyr, or the age of the Solar System. Thus, at each integration step and for each asteroid, a random number is generated between 0 and 1; if it is less than $\mathcal{P}$, the asteroid is destroyed. 

We note that this correction for the collisional loss assumes the current state of the Solar System for the entire simulation, and does not take into account the fact that the collisional loss was greater in the earlier days in the Solar System. The collisional loss should decrease as a function of time, as there are less bodies for other bodies to collide into. To compensate for this, \citet{Bottke2005Icar..175..111B} estimated that prolonging the total integration time by 10~Gyr in the current state of the Solar System is roughly equivalent to integrating for 4.5~Gyr in a Solar System that experienced higher collisional loss in its earlier days. We therefore perform two corrections, one compensating for dynamical loss over 4.5~Gyr and the other for 10~Gyr. 

In total, we simulate the collisional loss of the planetesimals 10,000 times. We are then able to find the average fractional loss as a function of diameter, which is the ratio, $\mathcal{F}_{\text{collision}}(D) = C_{\text{final}}(D)/C_{\text{initial}}(D)$, or the final number of counts of asteroids of size $D$ divided by the initial number of counts. 

At the same time, we compensate for the dynamical loss \citep[see the supplemental materials of][]{Delbo2017}: we search for the fractional loss as a function of diameter. To do this, we created a sample of asteroids from the inner main belt, with low inclination, and low eccentricity orbits, particularly those whose perihelion does not cross the orbit of Mars. Then, the simulation is evolved for 4.5~Gyr. The asteroids feel the effects of all the planets from Venus to Uranus, Ceres, and Vesta, and the Yarkovsky effect. The integration time step is 10~days. To compensate for such a high temporal resolution, only about 600~bodies are used for this integration. A body is removed when its semi-major axis falls below 0.5~au or beyond 30~au, or if it comes within the Hill Sphere of a planet. At the end of the simulation, the dynamical fractional loss, $\mathcal{F}_{\text{dynamic}}(D),$ is found using the same method as described above. 

Once we have the average collisional and dynamical loss fractions, we then estimate the original SFD simply by multiplying their fractions:
\begin{equation}
C_{\text{original}}(D) = \frac{1}{\mathcal{F}_{\text{dynamic}}} * \frac{1}{\mathcal{F}_{\text{collision}}} C_{\text{observed}}(D).
\end{equation}

\section{Results}
\label{S:Results}

\subsection{Synthesis of family removal}

\begin{figure*}
    \centering
    \includegraphics[width=\textwidth]{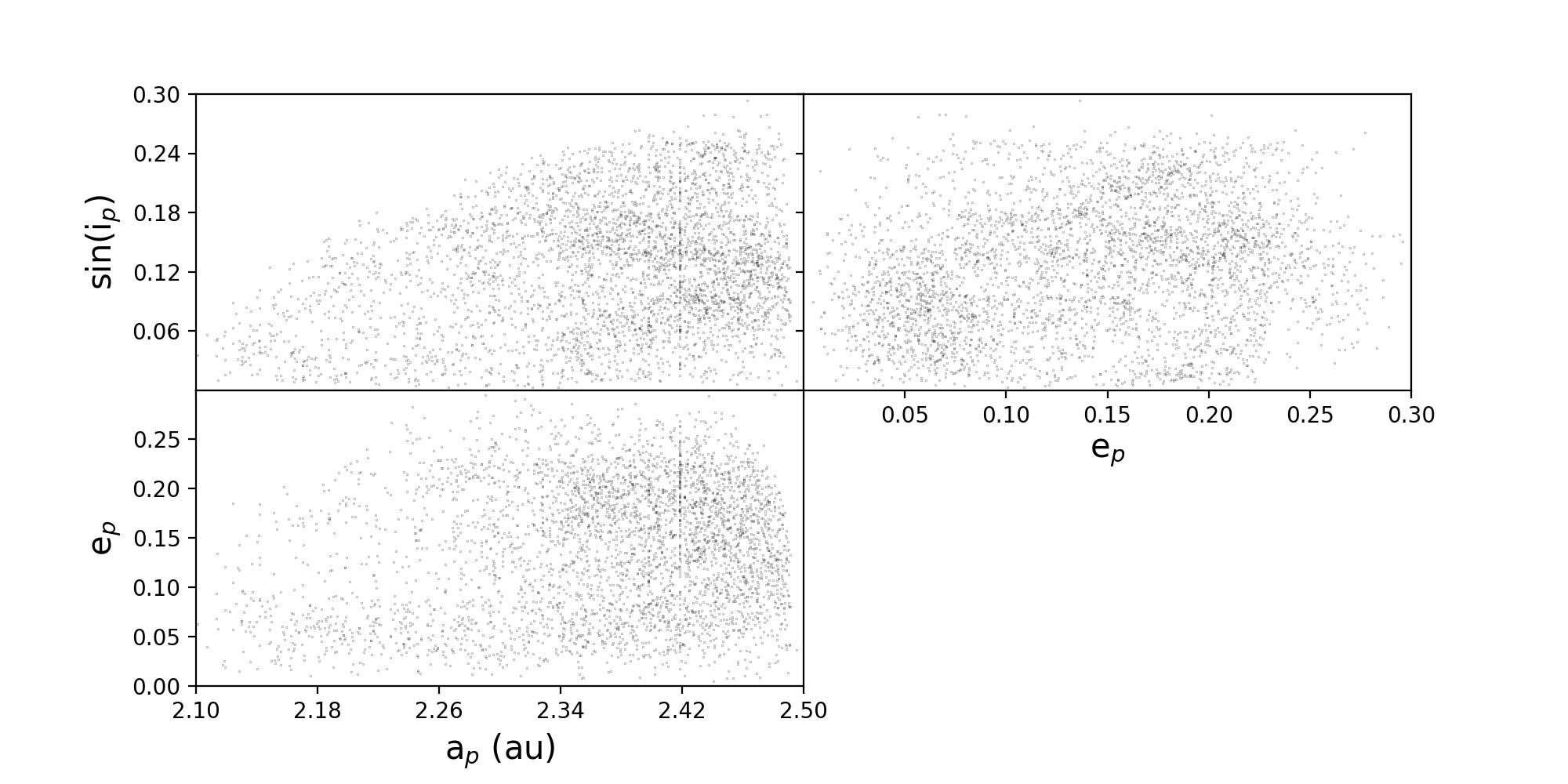}
    \caption{Inner main belt after removing all of the asteroid families identified in Table~\ref{tab:hcm-simulated-bg} given our membership reassessment.}
    \label{fig:proper-elements-no-families}
\end{figure*}

An example of the V-shape-constrained HCM method is presented in Fig.~\ref{fig:baptistina-vshape} for the case of the Baptistina family. This method was also applied to all the families labeled in Table~\ref{tab:hcm-simulated-bg}. The figure shows how the V-shape of the family is used to constrain the population of asteroids to which HCM clustering is applied. We note that this restriction applies to $a_p$ but not $e_p$ and $i_p$. For each of the plots of Fig.~\ref{fig:baptistina-vshape}, the $v_c$ is 150 $\textrm{m s}^{-1}$. We note that this central body, (298) Baptistina, is near the vertex of the V-shape and that the second body has a much smaller diameter and is therefore not adjacent to (298) Baptistina in this space. However, in proper element space, these bodies are adjacent. As expected, an asteroid family is recovered from the real data, yet not from the synthetic data, which have no clusters.

While viewing Fig.~\ref{fig:baptistina-vshape}, we notice in sin($i_p$) and $e_p$ space that a small satellite cluster located at (0.12,0.10) is included in the Baptistina family. This is part of the Vesta family and overlaps with the 
V-shape of Baptistina. Although a piece of the Vesta family is included in our Baptistina family, this is not problematic because our primary objective is to remove all known collisional family members. Specifying the family of a specific body is not a primary objective. Nevertheless, we present a supplementary analysis whereby we obtain a robust family membership using our V-shape-constrained HCM in Appendix~\ref{appendix:knownFamMembhersip}.

We estimate the uncertainty introduced by the free parameter N$_d$ by removing the families for each $v_c$ listed in Table~\ref{tab:hcm-simulated-bg}. In other words, we remove the families seven different ways for each density listed in Table~\ref{tab:hcm-simulated-bg}. By doing so, we are left with seven family-removed datasets, $\{a_{p},e_{p},i_{p},D\}_{\text{FR}}$, with a number of bodies of 3866, 3866, 3707, 2735, 1932, 2278, and 1235, for each N$_d$ fraction of 3/4, 2/3, 1/2, 1/3, 1/4, 1/5, and 1/10, respectively. We note that the case N$_d$=1/4 has less bodies than the case N$_d$=1/5. This is due to the value of $v_c$ found for Euterpe, which jumps to a large number from N$_d=1/3$ to N$_d=1/4$ because of our choice of using the median value and because the distribution of $v_c$ per family per N$_d$ may be under-sampled (see Fig.~\ref{fig:velocityCutOffHistogram}). We show the conclusions found in the following steps are valid even when considering all of the tested N$_d$. 

One of the family-removed datasets is presented in Fig.~\ref{fig:proper-elements-no-families}. This is for the $v_c$-value found for using either one of the following values for the free parameter N$_d=[1, 3/4, 2/3]$, since the results are equivalent. 
As expected, there are no obvious signs of other families as clusters in proper-element space. Importantly, we do not see obvious signs of family ``halos'', which, on the other hand, are clearly visible if we remove families according to a more conservative membership, as in Fig.~\ref{fig:backgroundNesvorny}.

\subsection{Detection of an ancient family and planetesimal identification}
\label{detectionofanancientfamily}
\begin{figure*}
    \centering
    \includegraphics[width=0.31\linewidth]{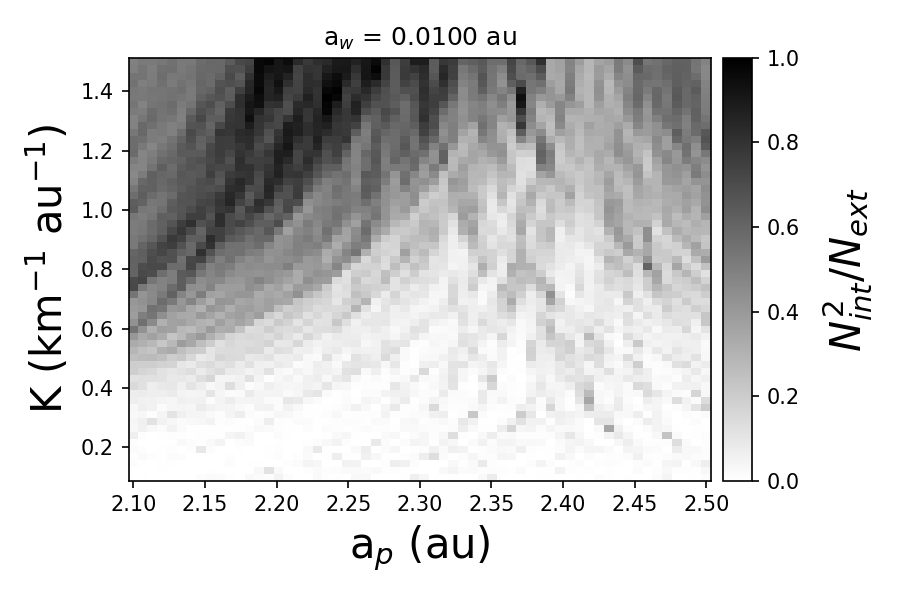}
    \includegraphics[width=0.31\linewidth]{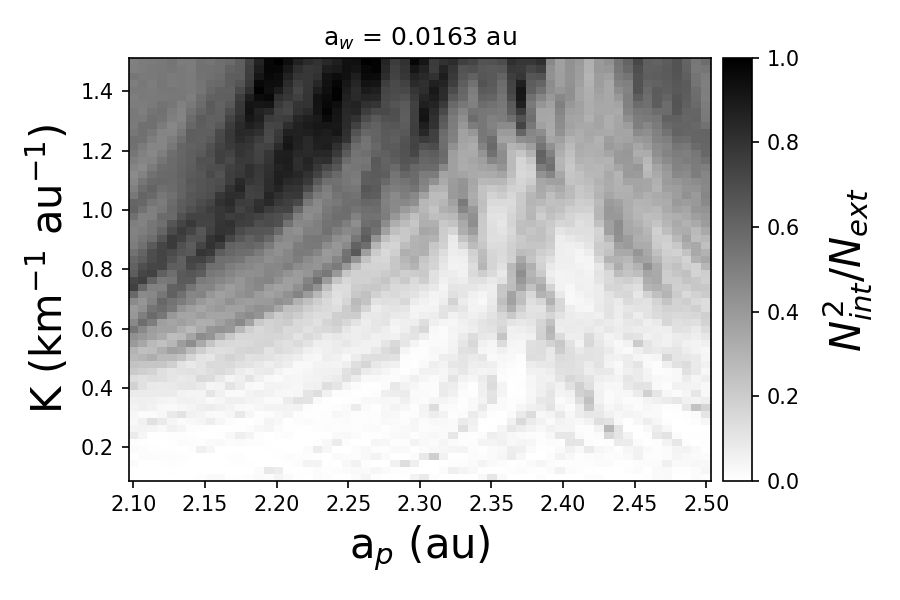}
    \includegraphics[width=0.31\linewidth]{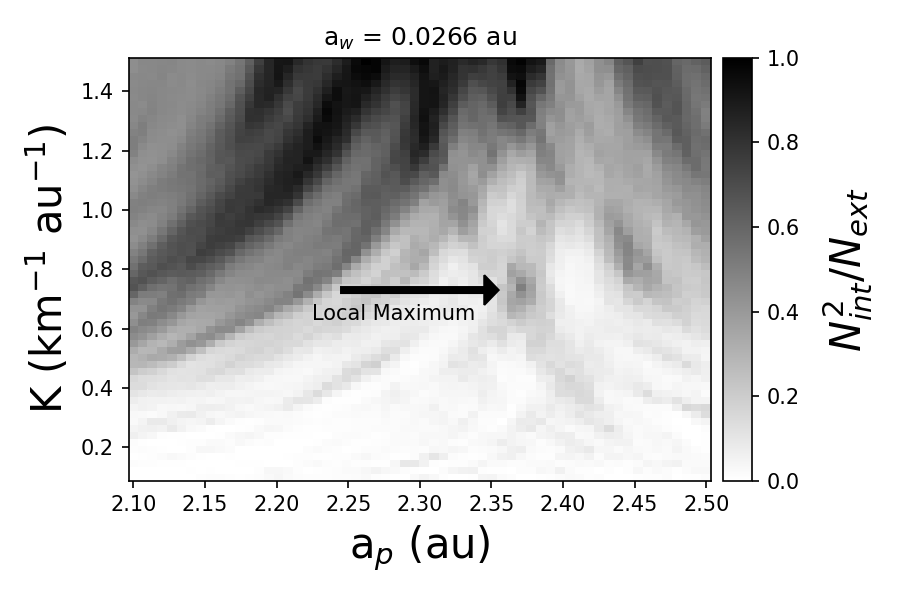}

    \includegraphics[width=0.31\linewidth]{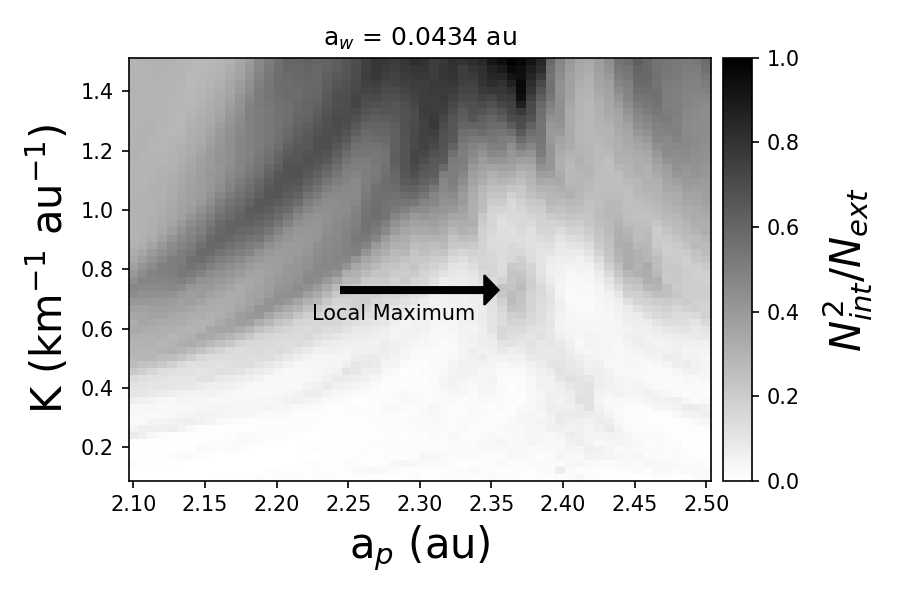}
    \includegraphics[width=0.31\linewidth]{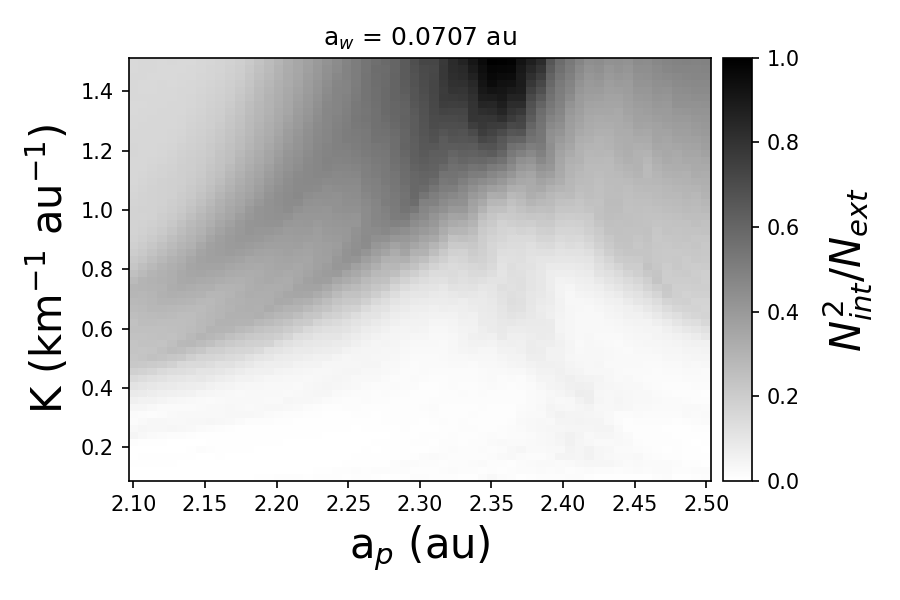}
    \includegraphics[width=0.31\linewidth]{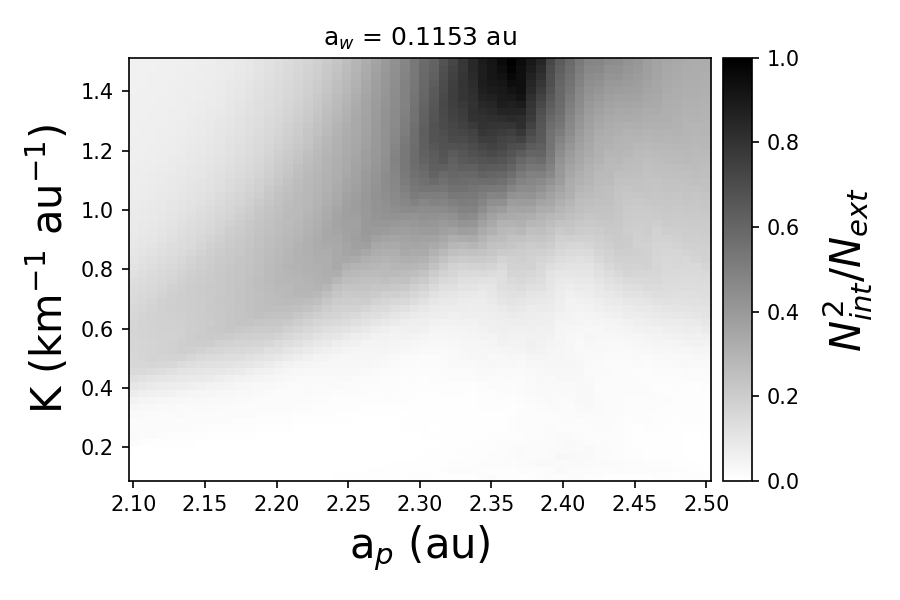}

    \includegraphics[width=0.31\linewidth]{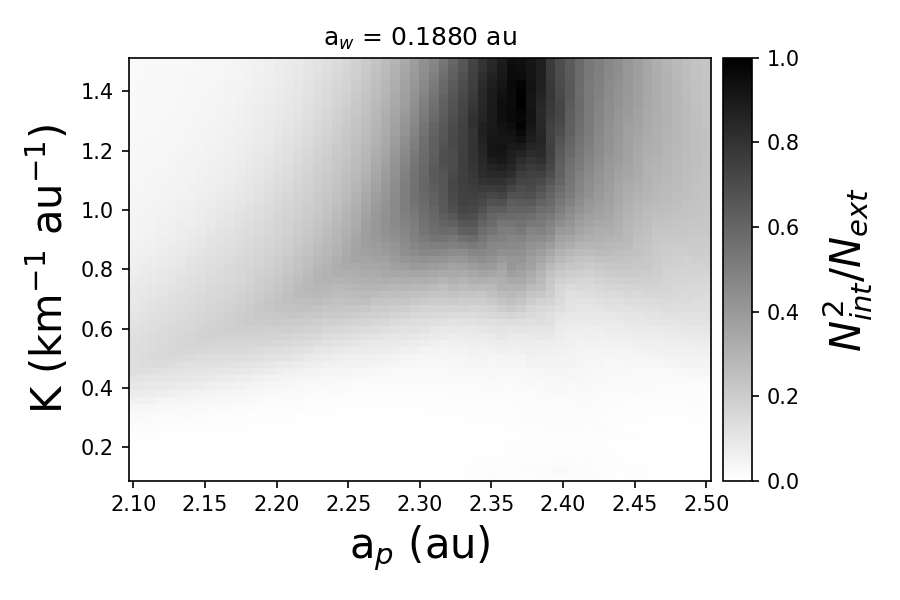}
    \includegraphics[width=0.31\linewidth]{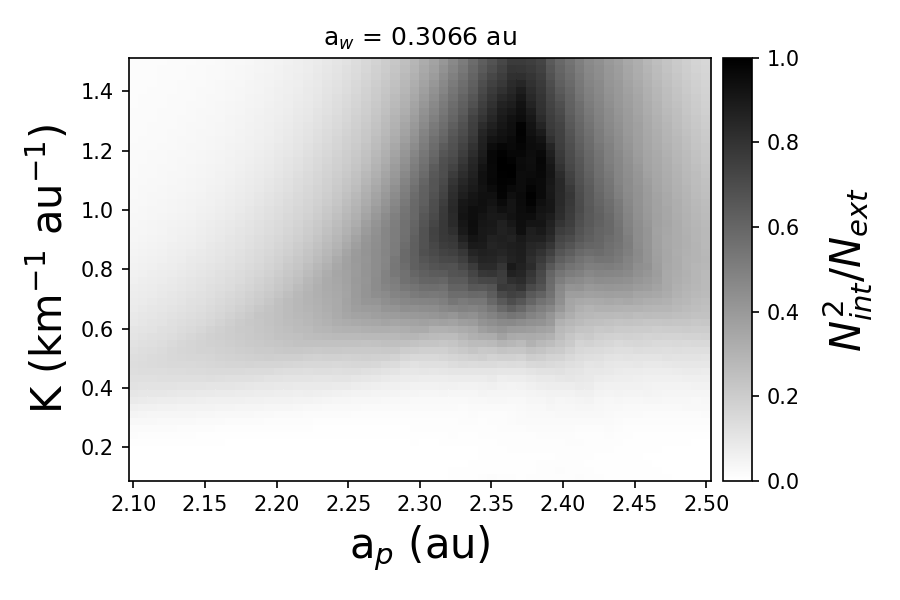}
    \includegraphics[width=0.31\linewidth]{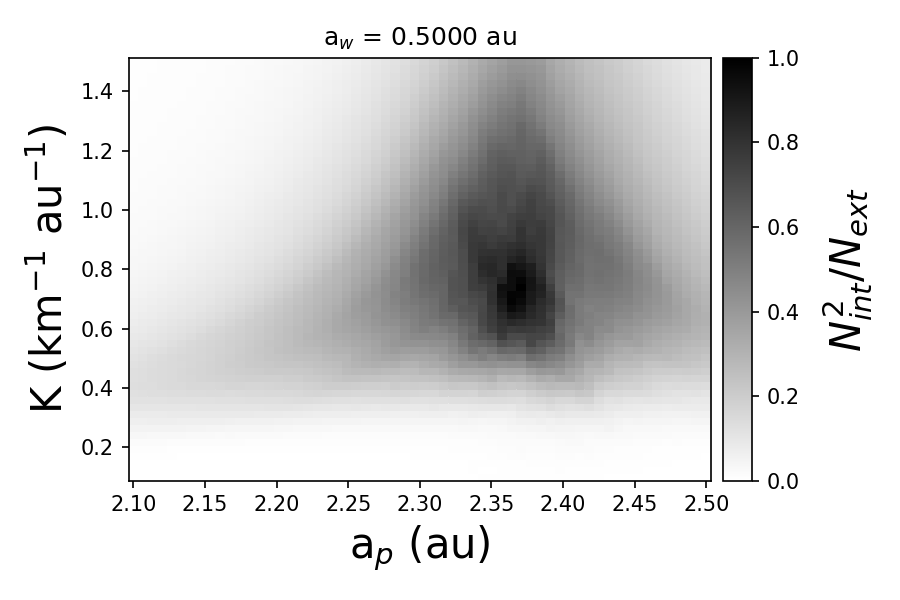}
    
    \caption{Each score map is a stacked average of 140 score maps comprising 20 realizations of the diameters for each of the seven choices of the $v_c$ corresponding to N$_d$. Each score map uses a different lobe width, as indicated by the respective titles. The scores are stretched and normalized such that all values are between zero and one. The signature of the local maximum indicating the ancient family is labeled for two lobe widths.}
    \label{fig:global-search}
\end{figure*}

Having removed previously known families, we perform the global search for undetected families. As there are
no obvious signs of clusters in the family-removed dataset of Fig. 6, we turn to the V-shape search. We consider nine different $a_w$-values, taking into account the effects of the free parameter N$_d$ and the uncertainty on asteroid diameters. First, in the score maps shown in Fig.~\ref{fig:global-search}, we can see that the map created with $a_w$=0.01~au is too sporadic and the global maximum is a random ``hot pixel'' that does not correspond to a large-scale trend indicative of the presence of a family, but is rather characterized by stochastic sampling. Next, considering $a_w$=0.0707~au and upward, we see a global maximum that remains invariable within the range 2.35--2.40~au. This is the expected behavior for a detection  of a family V-shape. The behavior along the $K$-axis is more complex. As demonstrated by \citet{Deienno2021}, we notice the trend that the global maximum on the score map for a family detection decreases in slope as the width, $a_w$, increases. We notice a local maximum at $(a_c, K)$ = (2.36 au, 0.7~${\text{km}^{-1}\text{au}^{-1}}$) that comes into focus and eventually fades away for $a_w$ between 0.0266 and 0.434~au. This local maximum now represents our candidate family. 

\begin{figure}
    \centering
    \includegraphics[width=0.99\linewidth]{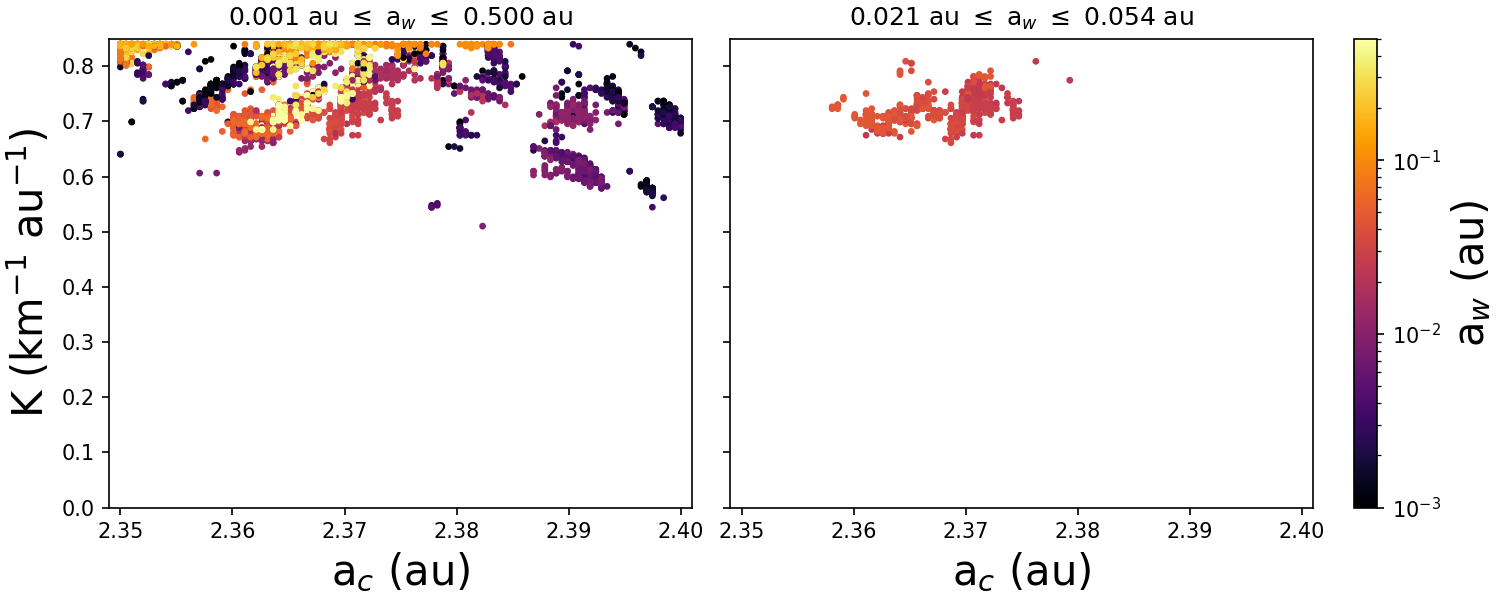}
    \caption{Demonstration of how the widths of the $a_w$ method were selected. The color indicates the width of the lobes. Each point shows the best found slope and vertex. In the left panel, there are 4,200 points (seven values of the $v_c$, 20 iterations of the diameters uncertainty, and 30 lobe widths.) The right panel is a filtered version of the left, only including lobe widths that converge stably to a solution and includes 700 points.}
    \label{fig:establishing_good_width}
\end{figure}

Next, we aim to fine tune this detection by restricting the range of slopes and vertexes so that the local maximum over this domain is also the global maximum. After the scan range is reduced, we eventually study the behavior of the best $a_c$ and $K$---which are coordinates of the maxima on the local score maps---in relation to $a_w$. We reduce $a_c$ to 2.35 and 2.40~au. We reduce the slope range to scan between 0 and 0.84~${\text{km}^{-1}\text{au}^{-1}}$. A 75$\times$75 linearly spaced grid is created within these limits. We then use a higher resolution of 30 values of $a_w$  logarithmically spaced between 0.001 and 0.5~au. Again, in this procedure we take into consideration the dependence of the results on the  uncertainty in diameter. Thus, we create 4,200 score maps, that is, one for each combination of the seven N$_d$, the 20 realizations of the diameters, and the 30 values of $a_w$.  Figure~\ref{fig:establishing_good_width} is a compilation of these results; contrary to Fig.~\ref{fig:global-search}, which shows an average score map per $a_w$, the left panel of Fig.~\ref{fig:establishing_good_width} reports only the maximum found per score map and each marker is color coded according to $a_w$. Then, the right panel shows the filtered and accepted values of $a_w$. 

To filter the range of $a_w$, we clip a lower and upper limit of the left panel of  Fig.~\ref{fig:establishing_good_width} until the resultant score-maxima converge. When $a_w$ is less than 0.021~au, we notice that the values are rather sporadic and maxima span the entire $a_c$ domain. In regard to the upper limit, we see that $K$ correlates too strongly with $a_w$. For instance, between 0.054 and 0.4~au, the maximum scores are found near to the upper border of the search range. Thus, $a_w$ between 0.021 and 0.054~au is the optimal range because the solutions converge. We note that the largest values of $a_w$ also converge to the same solution. We note that the largest values of $a_w$ also converge to the same solution. However, we opt not to use this for two reasons, the first being that $K$ still correlates too strongly with $a_w$. This is not the case for the optimal range. Secondly, we use the lobes to determine membership, and large lobes will include more bodies that may be interlopers.

\begin{figure}
    \centering
    \includegraphics[width=\linewidth]{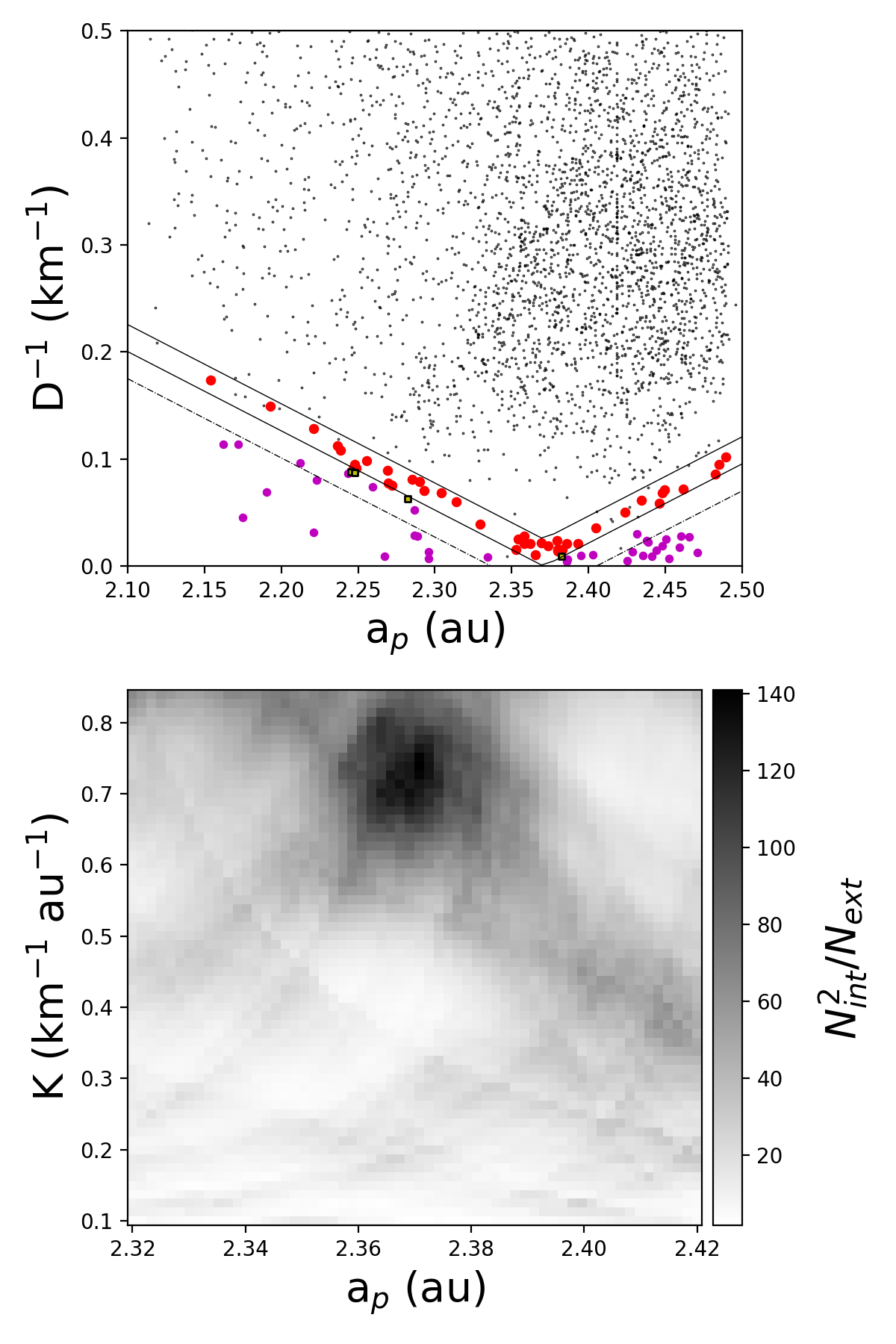}
    \caption{The detection of an ancient family through the Yarkovsky V-shape. \textit{Top:} The V-shape down in the $a_p,1/D$ plane. The inner and outer lobes of the V-shape are shown; these are requisite of the detection method, as described in the main text. As a proxy for membership, those within the interior lobes are considered to be family members and are shown as red dots. The purple dots are planetesimals, gray dots are uncategorized, and squares could be either planetesimals or ancient family members. \textit{Bottom:} Score map consisting of a stacked average of 700 Monte Carlo iterations in detecting the ancient family. The maxima gives the coordinates of the V-shapes vertex and slope parameters, $a_c$ and $K$.}
    \label{fig:ancient-family-detection}
\end{figure}

By fine-tuning our detection, we retain five $a_w$-values tested from 0.021 to 0.056~au. This, in tandem with the 20 realizations of the diameters and the 7 family-removed datasets, leads us to a total number of 700 score maps, which are stacked together and presented in the bottom panel of Fig.~\ref{fig:ancient-family-detection}. In regard to the top panel, we show the nominal V-shape of the ancient family, which is the middle V. The inner and outer Vs are the lines constraining the lobes, using the mean of the filtered $a_w$ values. Next, we color the planetesimals in purple, the ancient family members in red, and those that could be either in yellow as squares with black edges. 

To determine which category a body belongs to, we track the number of times a body falls within the inner lobe and those that are below the determined V-shape for each of the 700 iterations. Those inside the lobe for more than 1$\sigma$ of all iterations are nominal family members, which represent only a subset of the total family members, namely  those experiencing maximum Yarkovsky drift. Those that are below the V-shape are planetesimals. Those bodies that are in yellow belong to either group for more than 68\% of the iterations. The rest of the bodies above in gray we now consider to be uncategorized because they could be either members of the ancient family, missed members of the known families, or members of undetected families; they could also be planetesimals, but this is less likely because, as we demonstrate in section~\ref{S:Discussion}, their SFD is more consistent with those of families of fragments than with those of the planetesimals. We note that some bodies fall within the interior lobe and are colored black instead of red; this is in light of the presented realization of their diameters in the Monte Carlo sampling; that is, they fell into the lobe for this iteration but not more than 68\% of all iterations. In Appendix~\ref{appendix:ancient-primordial-members}, we report a list of nominal family members and planetesimals. Additionally, we show the proper orbital elements in Fig.~\ref{fig:proper-elements-ancient-family}. Unsurprisingly, the members do not clump but are rather disperse.

With the slope and the vertex of the family's V-shape calibrated, we now follow the procedure identified in step 3 of section~\ref{sec:vsearch} to establish the detection confidence. Briefly, we find the ratio of bodies interior to the V to those exterior for 700 different iterations of the nominal data and 70,000 different iterations of the shuffled dataset; that is, 100 shufflings of the semi-major axis for each of the realizations of the nominal data. In all, we find 43/70,000 iterations, where $R^{shuf}_{in/out} > R^{nom}_{in/out}$, leading to a statistical significance of 3.4$\sigma$. With the detection confidence established, we return our attention to the right panel of Fig.~\ref{fig:establishing_good_width}, where we find the mean $K$ and $a_c$ value as well as the cross correlation. We find that the V-shape has a slope of $0.73\pm 0.03$~km$^{-1}$~au$^{-1}$, a$_c$ has a best value of $ 2.369 \pm 0.004$~au, and the two variables have a covariance, $\sigma^{2}_{\text{K},\text{a}_p}$, of 0.0077~km$^{-2}$ , which is a correlation of 0.55.

\subsection{The corrected size--frequency distribution}
\label{S:sfdcorrected}

\begin{figure}
    \centering
    \includegraphics[width=.99\linewidth]{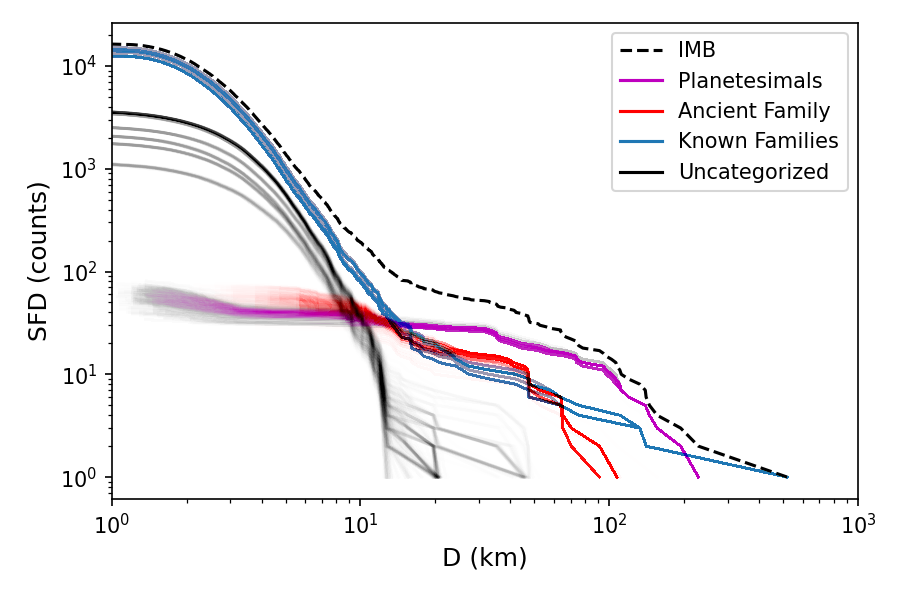}
    
    \includegraphics[width=.99\linewidth]{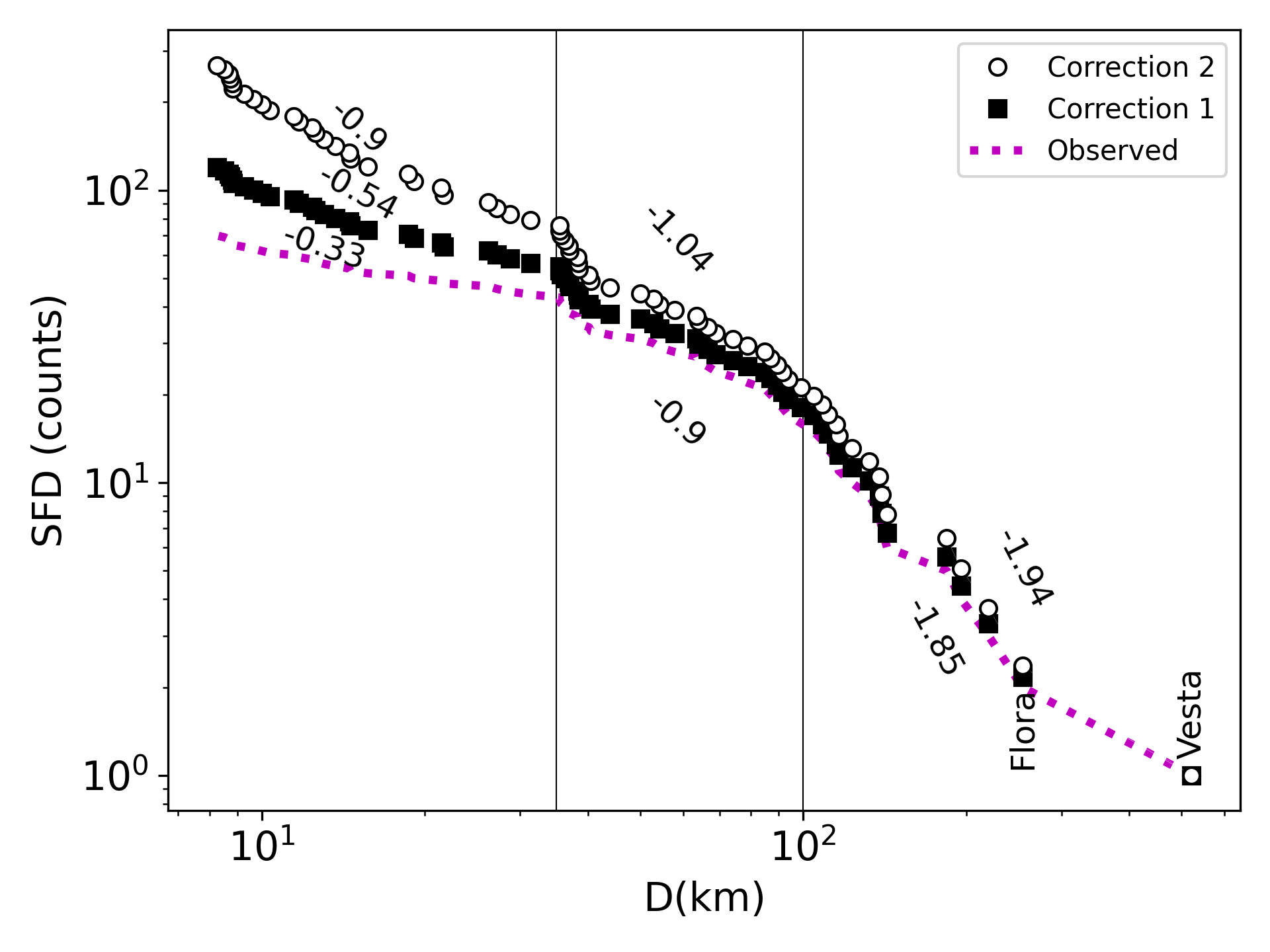}

    \caption{SFDs of various collections of asteroids. \textit{Top:} Cumulative SFD of five different cases. The high-$p_{V}$ inner main belt is indicated as IMB, the blue Known Families are the reassessed families listed in Table~\ref{tab:hcm-simulated-bg}. The {uncategorized} objects are those that do not belong to the known families, the newly discovered ancient family, or the planetesimals. Multiple curves are drawn according to each iteration in order to reflect the uncertainty. There are 700 highly transparent curves drawn for the Ancient Family, the Planetesimals, and the Uncategorized. There are seven known family curves drawn, and one curve for the inner main belt. \textit{Bottom:} Purple curve labeled ``{Observed''} is a modified version of the Planetesimals from the top panel. This list is a compilation of planetesimals detected with a greater than 1$\sigma$ confidence, identified planetesimals from previous works, and a few artificial asteroids that should be hidden behind the  V-shapes of the ancient families, as explained in the Appendix. Next, we provide two different corrections for depletion. Both consider the same dynamical loss while correction 2 accounts for more collision loss than correction 1, as explained in the text. We provide the power-law slope in three different regions: below 35 km, between 35 and 100 km, and above 100 km. We note that the three slopes are reported below 35 km, yet above we only report Correction~1~and~2 for clarity. }
    \label{fig:SDFs}
\end{figure}

With the detection of the ancient family established, we can now look at the SFDs of various subsets of the inner main belt, such as: the reassessed known families, the ancient family, the planetesimals, and the remaining uncategorized bodies. This is shown in the top panel of Fig.~\ref{fig:SDFs}, where the fuzziness of the lines corresponds to the uncertainties from each iteration in the Monte Carlo. Interestingly, the uncategorized bodies have a very similar shape to the known families, indicating that these objects are also collisional fragments. We note that the SFDs of the planetesimals have a much different shape from the SFDs of the known families. Beyond that, if we look at the SFD of the high $p_{V}$ population of  the entire inner main belt, we notice two features that are both humps: the planetesimals corresponding to the first hump of larger bodies on the right, and the known families characterizing the hump for smaller bodies on the left.

The SFD of the ancient family is unlike those of the known families and the planetesimals. However, this is expected, as we know the membership of the ancient family is incomplete, and many are uncategorized. Furthermore, there may be some planetesimal interlopes within the inner lobes, particularly the large objects that crowd the vertex of the V-shape. Removing only a few large-diameter objects would have a significant effect on the shape of the SFD, particularly in the cumulative count range of C<10.

To finish, we wish to explore the possible SFD of planetesimals in the inner main belt during the early days of the Solar System. To do this, we take the list of objects identified at the 1$\sigma$ confidence level from Table~\ref{tab:primordial-objects} and add the other known planetesimals from \citet{Delbo2017} and \citet{Delbo2019}, which totals  54 bodies. However, given that the planetesimals were identified through a process of elimination, that is, they were not within any family's V-shape in the $a$--$1/D$ plane, and that the size and semi-major axis should be uncorrelated for the planetesimals, we can expect more planetesimals to be hidden amongst the ancient family's V-shapes. We therefore compensate for the V-shape obscuration, as explained in Appendix~\ref{appendix:vcomp}. Briefly, this compensation is treated in a Monte-Carlo fashion; we often generate between 12 and 25 additional bodies. In general, these bodies have small diameters, which is expected because the magnitude of the obscuration is inversely proportional to the diameter. As a result, the compensation does not change the reported slope for the SFD for bodies larger than 100 km, yet increases the reported slopes by a factor of 1.4 for objects smaller than 30 km and is therefore not negligible.

Next, as described in Section~\ref{sec:methods}, we correct for dynamical and collisional loss. Two models are presented, both experience comminution rates based on the current state of the inner main belt. One compensates for this collisional loss over 4.5 Gyr, which would be a lower limit given that the loss rate today is lower than before. The second applies the correction for 10 Gyr, which attempts to balance stronger comminution in the past. These in turn establish a possible range of SFDs we would expect for the original planetesimal population and are shown in the bottom panel of Fig.~\ref{fig:SDFs}.

\section{Conclusions and discussion}
\label{S:Discussion}

Given our new technique of applying the HCM on a family-by-family basis, each time restricted by the Yarkovsky V-shape of the family, we were able to reassess the membership of known asteroid families with high albedo, and the issue of ``halos''.  Subsequently, we removed these known families from the inner main belt and applied a V-shape search method to the remaining population: by doing so we identified one of the most ancient asteroid families, which has the widest open V-shape. From determining the slope and the vertex of the family's V-shape, we have two of the three necessary parameters for an age estimate. Finally, with the most ancient family identified, we are able to isolate planetesimals and present a range of possible SFDs given different corrections for dynamical and collisional loss.

In regards to the family's age, as shown in Eq.~\ref{eq:age-estimation}, its computation depends on the value we adopt for the bulk densities of the member asteroids. \citet{vernazza2021vlt} were able to constrain the shapes of many asteroids using imaging and spectro-polarimetry. If we take the 12 S-type asteroids reported, we obtain a bulk density of $3.0 \pm 0.3$ g~cm$^{-3}$, which corresponds to an age estimation of $5 \pm 1$~Gyr.  At the same time, the MP$^3$C database has masses and diameters reported for a handful of members, which are listed in the caption of Table~\ref{tab:ancient-family}. Assuming the asteroids are spheres, we obtain bulk density measurements of $2.6 \pm 1 $ g~cm$^{-3}$, which results in an age estimate of $4.3 \pm 1.7$~Gyr. However, it must be taken into account that the age estimation has large error bars. A numerical analysis of the dynamics of this family may be performed---similarly to the work presented in \citet{milani2017ages}---to reduce the uncertainty.

In either case, the age of this asteroid family is at the limit of the age of the Solar System. Finding a more precise measurement of the age could add a valuable constraint to Solar System evolution models, especially for what concerns the timing of the dynamical instability that affected the orbits of the giant planets. Moreover, the ancient family that we have found is so dispersed in eccentricity and inclination that it must predate the giant planet instability. The interplay between Yarkovsky effect and resonances cannot produce such a dispersed population; only the giant planet instability can \citep{brasil2016dynamical}; see \citet{Tsiganis2005Natur.435..459T} for the original concept of the giant planet instability.

With the detection of this ancient family, one low-albedo family from \citet{Delbo2017}, and two X-complex families from \citet{Delbo2019}, there is now a total of four asteroid families discovered within the inner main belt with ages greater than 2 Gyr. However, there are about 25 other families within the inner main belt whose ages are younger than 2 Gyr. As explained in \citet{Bottke2005Icar..175..111B}, the collisional environment is estimated to have been more violent in the early days of the Solar System and as a result we should expect to observe more ancient asteroid families than young families. However, this is not the case. Although pairing the V-shape with the HCM represents a significant advancement in asteroid family detection, the method still has its limitations and therefore detectable families may exist that are still beyond our current detection capabilities. As shown in \citet{Deienno2021}, detection efficiency drops greatly as a family's contrast decreases with respect to the background population.

On the other hand, recent studies have emerged suggesting that the asteroid belt was not necessarily as massive upon formation as it is today, which would be consistent with the finding that there are less ancient families than younger ones. Indeed, \citet{raymond2017empty} performed a numerical simulation demonstrating a viable mechanism of populating an originally empty belt that results in the observed S/C-type distribution as a consequence of the formation of terrestrial and giant planets. Other studies indicate that planetesimals formed in ring-like structures at certain distances about their host stars and as such must have been transported to other locations later \citep{morbidelli2022contemporary,izidoro2022planetesimal}. Additionally, \citet{deienno2022implications} studied the effects of giant planet instability and radial migration on the asteroid belt and found that a high initial solar system mass could not reproduce the low-mass belt that we observe today. Future numerical studies of family formation and dissipation in the two paradigms of either a massive or sparse initial asteroid belt are necessary in order to determine which is more likely to reproduce the observed distribution in collisional family ages observed today.

Given dynamical depletion, perhaps only ancient asteroid families with a high initial number density remain detectable today regardless of detection efficiency. However, to know if this depletion can compensate for the observed difference between the number of young and ancient families---of about a factor 5---requires full-scale dynamical evolution simulations of family generation with collisional models and detection. It is possible that such an experiment will show that dynamical depletion cannot explain the observed discrepancy. In such a case, the estimated collisional rates of the early Solar System may need to be revisited.

The SFDs shown in Fig.~\ref{fig:SDFs} have important implications. To contextualize the result within the literature, we must note that the presented SFD considers a new compensation that previous works did not; namely we proportionally increase the count of observed bodies to include those that are potentially obscured by the V-shapes of ancient families. With this in mind and with the discovery of this new ancient family, we have been able to identify more planetesimals than previous works, particularly bodies smaller than 50 km in diameter \citep{Delbo2017}. Notably, the SFD of the low-albedo inner-main-belt planetesimals identified by \citet{Delbo2017} was hindered by lacking bodies smaller than 35 km in diameter. To combat this, the authors extrapolated a power-law slope for $D$ $<$ 35~km. Their predictions, when adjusted for the V-shape obscuration, are consistent with the power law deduced from our analyses, which now contains $D$ $<$ 35~km planetesimals. 

Second, the SFD of the ``uncategorized'' objects of Fig.~\ref{fig:SDFs} resembles the SFD of the known families, supporting the conclusion that they too are collisional fragments and less likely to be original planetesimals. Before carrying out our analysis, we expected uncategorized bodies to be collisional fragments originating from either the known asteroid families, the newly detected ancient asteroid family, the undetected asteroid families, or the planetesimals. Another a priori expectation was that the SFD of the planetesimals may be incomplete, since some of its members may be amidst the uncategorized objects. However, as the SFD of the uncategorized objects indicates that they are collisional fragments, we infer that the deduced planetesimal SFD is complete or at the very least not missing a major component. 

Thirdly, the inferred original SFD of the planetesimals can be used for Solar System evolution models \citep{klahr2022}. Interestingly, the planetesimal SFD has the steepest power-law index in the regime of $D > 100$~km, a turning point around $D \sim 100$~km,  a shallower power index for $35 < D < 100$~km,  and an even shallower power index for $D<35$~km, which indicates that planetesimals mostly formed as large objects. This result is consistent with previous studies of the original asteroid population in the inner main belt \citep{Delbo2017,Delbo2019}, in other regions of the main belt \citep{tsirvoulis2018reconstructing}, and with numerical simulations that searched for the original SFD of planetesimals that produced---by collisional objects---an SFD that is consistent with the current SFD of the asteroid belt \citep{Bottke2005Icar..175..111B,Morbidelli2009Icar..204..558M}. 

Our results regarding the  SFD of the original planetesimals are also consistent with theoretical studies of planetesimal formation by pebble cloud collapse in turbulent protoplanetary disks. In particular, \citet{klahr2020turbulence} predict a relatively narrow differential size distribution of planetesimals with a best effective size of around 100 km; this can be intuitively explained by the fact that smaller pebble clouds---with $D \ll 100$~km---are unlikely to collapse into planetesimals, because their self-gravity is overcome by turbulent diffusion, while and larger pebble clouds---with $D \gg 100$~km---would take longer to accumulate than the time they would need to collapse \citep[see also][for a review]{klahr2022}.

Lastly, the list of bodies we identify as planetesimals is robust against the free parameters and uncertainties from the reassessment process of the known families. Potential false positives among the identified primordial objects may be fragments from asteroid families that have disassociated beyond a detectable limit or fragments from families that formed before the giant-planet instability, which may have erased the dynamical linkage between members. However, given that the results are insensitive to the rigour with which the family-removal procedure is applied, we expect this number to be a small contribution. Moreover, the robustness supports that the planetesimals and collisional families are indeed two independent populations. That is, the planetesimals are processed through collisional evolution and create the collisional asteroid families, which is a conclusion also supported by \citet{Delbo2017}.


A series of follow-up studies could be carried out to either complement this work, benefit from the newly developed methods, or exploit its results. Firstly, the membership lists provided for either the ancient family or the planetesimals contain some interlopers (false positives). For example, knowing the spin states of the asteroids and obtaining their visible and near-infrared spectra could help to identify these interlopers \citep[e.g.,][used the spin states to do just this]{athanasopoulos2022asteroid}. A plethora of visible and near-infrared spectra already exist \citep{usui2011asteroid}, and with the new Gaia Data Release 3, the spin states for many asteroids are now available \citep{collaboration2022gaia}. Also, with more information about the ancient family, a better age estimate may be obtained, which could add a more robust constraint in planetary formation models.

Next, this method could be replicated for the high- and low-$p_{V}$ populations of the other sections of the main belt. After a reassessment and removal of other known families, we suspect that more planetesimals and ancient families will be discovered. Again, the discovery of the oldest families in other parts of the belt would provide temporal constraints to planetary formation models. Additionally, it would be interesting to see if the SFDs of the planetesimals in other sections of the belt have different shapes to those observed here. Though not the primary goal of this experiment, reassessment of the known families could be studied in further detail. A spectral analysis could be performed to compare the newly identified members in the ``halo'' with those in the core. At the same time, another analysis could attempt to disentangle the Baptistina and Flora families, which is not attempted here. 
Finally, we are hopeful that the SFD of the planetesimals can be used as a constraint in Solar System evolution models.

\begin{acknowledgements}
We acknowledge support from ANR “ORIGINS” (ANR-18-CE31-0014). This work is based on data provided by the Minor Planet Physical Properties Catalogue (MP$^3$C) of the Observatoire de la Côte d'Azur. 

KJW acknowledges support through Project ESPRESSO, a NASA SSERVI program at SwRI.

R.D. was supported by the NASA Emerging Worlds program, grant 80NSSC21K0387. 

\end{acknowledgements}

\bibliographystyle{aa} 
\bibliography{biblio}

\begin{thebibliography}{54}
\expandafter\ifx\csname natexlab\endcsname\relax\def\natexlab#1{#1}\fi

\bibitem[{Athanasopoulos {et~al.}(2022)Athanasopoulos, Hanu{\v{s}}, Avdellidou,
  Bonamico, Delbo, Conjat, Ferrero, Gazeas, Rivet, Sioulas,
  {et~al.}}]{athanasopoulos2022asteroid}
Athanasopoulos, D., Hanu{\v{s}}, J., Avdellidou, C., {et~al.} 2022, Astronomy
  \& Astrophysics, 666, A116

\bibitem[{{Birnstiel} {et~al.}(2016){Birnstiel}, {Fang}, \&
  {Johansen}}]{Birnstiel2016SSRv..205...41B}
{Birnstiel}, T., {Fang}, M., \& {Johansen}, A. 2016, \ssr, 205, 41

\bibitem[{Bolin {et~al.}(2017)Bolin, Delbo, Morbidelli, \& Walsh}]{Bolin2017}
Bolin, B.~T., Delbo, M., Morbidelli, A., \& Walsh, K.~J. 2017, Icarus, 282, 290

\bibitem[{{Bottke} {et~al.}(2006){Bottke}, {Vokrouhlick{\'y}}, {Rubincam}, \&
  {Nesvorn{\'y}}}]{Bottke2006}
{Bottke}, William~F., J., {Vokrouhlick{\'y}}, D., {Rubincam}, D.~P., \&
  {Nesvorn{\'y}}, D. 2006, Annual Review of Earth and Planetary Sciences, 34,
  157

\bibitem[{{Bottke} {et~al.}(2005){Bottke}, {Durda}, {Nesvorn{\'y}}, {Jedicke},
  {Morbidelli}, {Vokrouhlick{\'y}}, \& {Levison}}]{Bottke2005Icar..175..111B}
{Bottke}, W.~F., {Durda}, D.~D., {Nesvorn{\'y}}, D., {et~al.} 2005, \icarus,
  175, 111

\bibitem[{Bourdelle-de Micas {et~al.}(2022)Bourdelle-de Micas, Fornasier,
  Avdellidou, Delbo, Van~Belle, Ochner, Grundy, \&
  Moskovitz}]{de2022composition}
Bourdelle-de Micas, J., Fornasier, S., Avdellidou, C., {et~al.} 2022, Astronomy
  \& Astrophysics, 665, A83

\bibitem[{Brasil {et~al.}(2016)Brasil, Roig, Nesvorn{\`y}, Carruba, Aljbaae, \&
  Huaman}]{brasil2016dynamical}
Brasil, P., Roig, F., Nesvorn{\`y}, D., {et~al.} 2016, Icarus, 266, 142

\bibitem[{Bro{\v{z}} {et~al.}(2013)Bro{\v{z}}, Morbidelli, Bottke, Rozehnal,
  Vokrouhlick{\`y}, \& Nesvorn{\`y}}]{brovz2013constraining}
Bro{\v{z}}, M., Morbidelli, A., Bottke, W., {et~al.} 2013, Astronomy \&
  Astrophysics, 551, A117

\bibitem[{Brož \& Morbidelli(2013)}]{broz_eos_2013}
Brož, M. \& Morbidelli, A. 2013, Icarus, 223, 844

\bibitem[{Cuzzi {et~al.}(2008)Cuzzi, Hogan, \& Shariff}]{cuzzi2008toward}
Cuzzi, J.~N., Hogan, R.~C., \& Shariff, K. 2008, The Astrophysical Journal,
  687, 1432

\bibitem[{Deienno {et~al.}(2022)Deienno, Izidoro, Morbidelli, Nesvorn{\`y}, \&
  Bottke}]{deienno2022implications}
Deienno, R., Izidoro, A., Morbidelli, A., Nesvorn{\`y}, D., \& Bottke, W.~F.
  2022, The Astrophysical Journal Letters, 936, L24

\bibitem[{Deienno {et~al.}(2021)Deienno, Walsh, \& Delbo}]{Deienno2021}
Deienno, R., Walsh, K.~J., \& Delbo, M. 2021, Icarus, 357, 114218

\bibitem[{Delbo {et~al.}(2022)Delbo, Avdellidou, Bruot, \&
  Erard}]{delbo2022EPSC...16..323D}
Delbo, M., Avdellidou, C., Bruot, N., \& Erard, S. 2022, in Europlanet Science
  Congress 2022

\bibitem[{Delbo {et~al.}(2019)Delbo, Avdellidou, \& Morbidelli}]{Delbo2019}
Delbo, M., Avdellidou, C., \& Morbidelli, A. 2019, Astronomy \& Astrophysics,
  624, 1

\bibitem[{Delbo {et~al.}(2015)Delbo, Mueller, Emery, Rozitis, \&
  Capria}]{delbo2015asteroid}
Delbo, M., Mueller, M., Emery, J.~P., Rozitis, B., \& Capria, M.~T. 2015,
  Asteroid thermophysical modeling (University of Arizona Press Tucson,
  Arizona)

\bibitem[{Delbo {et~al.}(2017)Delbo, Walsh, Bolin, Avdellidou, \&
  Morbidelli}]{Delbo2017}
Delbo, M., Walsh, K., Bolin, B., Avdellidou, C., \& Morbidelli, A. 2017,
  Science, 357, 1026

\bibitem[{{Dermott} {et~al.}(2018){Dermott}, {Christou}, {Li}, {Kehoe}, \&
  {Robinson}}]{Dermott2018NatAs...2..549D}
{Dermott}, S.~F., {Christou}, A.~A., {Li}, D., {Kehoe}, T. J.~J., \&
  {Robinson}, J.~M. 2018, Nature Astronomy, 2, 549

\bibitem[{{Dykhuis} \& {Greenberg}(2015)}]{Dykhuis2015Icar..252..199D}
{Dykhuis}, M.~J. \& {Greenberg}, R. 2015, \icarus, 252, 199

\bibitem[{Emsenhuber {et~al.}(2021)Emsenhuber, Mordasini, Burn, Alibert, Benz,
  \& Asphaug}]{emsenhuber2021new}
Emsenhuber, A., Mordasini, C., Burn, R., {et~al.} 2021, Astronomy \&
  Astrophysics, 656, A70

\bibitem[{{Fornasier} {et~al.}(2016){Fornasier}, {Lantz}, {Perna}, {Campins},
  {Barucci}, \& {Nesvorny}}]{Fornasier2016Icar..269....1F}
{Fornasier}, S., {Lantz}, C., {Perna}, D., {et~al.} 2016, \icarus, 269, 1

\bibitem[{Galluccio {et~al.}(2022)Galluccio, Delbo, De~Angeli, Pauwels, Tanga,
  Mignard, Cellino, Brown, Muinonen, {et~al.}}]{collaboration2022gaia}
Galluccio, L., Delbo, M., De~Angeli, F., {et~al.} 2022, Astronomy \&
  Astrophysics

\bibitem[{Hanu{\v{s}} {et~al.}(2017)Hanu{\v{s}}, Viikinkoski, Marchis,
  {\v{D}}urech, Kaasalainen, Delbo, Herald, Frappa, Hayamizu, Kerr,
  {et~al.}}]{hanuvs2017volumes}
Hanu{\v{s}}, J., Viikinkoski, M., Marchis, F., {et~al.} 2017, Astronomy \&
  Astrophysics, 601, A114

\bibitem[{Izidoro {et~al.}(2022)Izidoro, Dasgupta, Raymond, Deienno, Bitsch, \&
  Isella}]{izidoro2022planetesimal}
Izidoro, A., Dasgupta, R., Raymond, S.~N., {et~al.} 2022, Nature Astronomy, 6,
  357

\bibitem[{{Johansen} {et~al.}(2007){Johansen}, {Oishi}, {Mac Low}, {Klahr},
  {Henning}, \& {Youdin}}]{Johansen2007Natur.448.1022J}
{Johansen}, A., {Oishi}, J.~S., {Mac Low}, M.-M., {et~al.} 2007, \nat, 448,
  1022

\bibitem[{{Klahr} {et~al.}(2022){Klahr}, {Delbo}, \& {Gerbig}}]{klahr2022}
{Klahr}, H., {Delbo}, M., \& {Gerbig}, K. 2022, in Vesta and Ceres. Insights
  from the Dawn Mission for the Origin of the Solar System (NASA), 199

\bibitem[{Klahr \& Schreiber(2020)}]{klahr2020turbulence}
Klahr, H. \& Schreiber, A. 2020, The Astrophysical Journal, 901, 54

\bibitem[{{Kne{\v{z}}evi{\'c}}(2017)}]{Knezevic2017SerAJ.195....1K}
{Kne{\v{z}}evi{\'c}}, Z. 2017, Serbian Astronomical Journal, 194, 1

\bibitem[{Knezevic \& Milani(2012)}]{knezevic2012asteroids}
Knezevic, Z. \& Milani, A. 2012, IAU Joint Discussion, P18

\bibitem[{Mainzer {et~al.}(2019)Mainzer, Bauer, Cutri, Grav, Kramer, Masiero,
  Sonnett, \& Wright}]{mainzer2019neowise}
Mainzer, A.~K., Bauer, J.~M., Cutri, R.~M., {et~al.} 2019, NASA Planetary Data
  System

\bibitem[{Marchi {et~al.}(2006)Marchi, Paolicchi, Lazzarin, \&
  Magrin}]{marchi2006general}
Marchi, S., Paolicchi, P., Lazzarin, M., \& Magrin, S. 2006, The Astronomical
  Journal, 131, 1138

\bibitem[{Masiero {et~al.}(2015)Masiero, DeMeo, Kasuga, \&
  Parker}]{masiero2015asteroid}
Masiero, J.~R., DeMeo, F.~E., Kasuga, T., \& Parker, A.~H. 2015, Asteroids IV,
  32, 3

\bibitem[{Masiero {et~al.}(2013)Masiero, Mainzer, Bauer, Grav, Nugent, \&
  Stevenson}]{masiero2013asteroid}
Masiero, J.~R., Mainzer, A., Bauer, J., {et~al.} 2013, The Astrophysical
  Journal, 770, 7

\bibitem[{Masiero {et~al.}(2011)Masiero, Mainzer, Grav, Bauer, Cutri, Dailey,
  Eisenhardt, McMillan, Spahr, Skrutskie, {et~al.}}]{masiero2011main}
Masiero, J.~R., Mainzer, A., Grav, T., {et~al.} 2011, The Astrophysical
  Journal, 741, 68

\bibitem[{{Milani} {et~al.}(2014){Milani}, {Cellino}, {Kne{\v{z}}evi{\'c}},
  {Novakovi{\'c}}, {Spoto}, \& {Paolicchi}}]{Milani2014Icar..239...46M}
{Milani}, A., {Cellino}, A., {Kne{\v{z}}evi{\'c}}, Z., {et~al.} 2014, \icarus,
  239, 46

\bibitem[{Milani {et~al.}(2017)Milani, Kne{\v{z}}evi{\'c}, Spoto, Cellino,
  Novakovi{\'c}, \& Tsirvoulis}]{milani2017ages}
Milani, A., Kne{\v{z}}evi{\'c}, Z., Spoto, F., {et~al.} 2017, Icarus, 288, 240

\bibitem[{Morbidelli(2002)}]{morbidelli2002modern}
Morbidelli, A. 2002, Modern celestial mechanics: aspects of solar system
  dynamics (CRC Press)

\bibitem[{Morbidelli {et~al.}(2022)Morbidelli, Baillie, Batygin, Charnoz,
  Guillot, Rubie, \& Kleine}]{morbidelli2022contemporary}
Morbidelli, A., Baillie, K., Batygin, K., {et~al.} 2022, Nature Astronomy, 6,
  72

\bibitem[{{Morbidelli} {et~al.}(2009){Morbidelli}, {Bottke}, {Nesvorn{\'y}}, \&
  {Levison}}]{Morbidelli2009Icar..204..558M}
{Morbidelli}, A., {Bottke}, W.~F., {Nesvorn{\'y}}, D., \& {Levison}, H.~F.
  2009, \icarus, 204, 558

\bibitem[{Nesvorn{\`y} {et~al.}(2015)Nesvorn{\`y}, Bro{\v{z}}, Carruba,
  {et~al.}}]{nesvorny2015identification}
Nesvorn{\`y}, D., Bro{\v{z}}, M., Carruba, V., {et~al.} 2015, Asteroids IV, 29,
  7

\bibitem[{Oszkiewicz {et~al.}(2015)Oszkiewicz, Kankiewicz, W{\l}odarczyk, \&
  Kryszczy{\'n}ska}]{oszkiewicz2015differentiation}
Oszkiewicz, D., Kankiewicz, P., W{\l}odarczyk, I., \& Kryszczy{\'n}ska, A.
  2015, Astronomy \& Astrophysics, 584, A18

\bibitem[{{Parker} {et~al.}(2008){Parker}, {Ivezi{\'c}}, {Juri{\'c}}, {Lupton},
  {Sekora}, \& {Kowalski}}]{Parker2008Icar..198..138P}
{Parker}, A., {Ivezi{\'c}}, {\v{Z}}., {Juri{\'c}}, M., {et~al.} 2008, \icarus,
  198, 138

\bibitem[{Raymond \& Izidoro(2017)}]{raymond2017empty}
Raymond, S.~N. \& Izidoro, A. 2017, Science advances, 3, e1701138

\bibitem[{Ryan \& Woodward(2010)}]{ryan2010rectified}
Ryan, E.~L. \& Woodward, C.~E. 2010, The Astronomical Journal, 140, 933

\bibitem[{Spoto {et~al.}(2015)Spoto, Milani, \&
  Kne{\v{z}}evi{\'c}}]{spoto2015asteroid}
Spoto, F., Milani, A., \& Kne{\v{z}}evi{\'c}, Z. 2015, Icarus, 257, 275

\bibitem[{Tedesco {et~al.}(2002)Tedesco, Noah, Noah, \&
  Price}]{tedesco2002supplemental}
Tedesco, E.~F., Noah, P.~V., Noah, M., \& Price, S.~D. 2002, The Astronomical
  Journal, 123, 1056

\bibitem[{{Tsiganis} {et~al.}(2005){Tsiganis}, {Gomes}, {Morbidelli}, \&
  {Levison}}]{Tsiganis2005Natur.435..459T}
{Tsiganis}, K., {Gomes}, R., {Morbidelli}, A., \& {Levison}, H.~F. 2005, \nat,
  435, 459

\bibitem[{Tsirvoulis {et~al.}(2018)Tsirvoulis, Morbidelli, Delbo, \&
  Tsiganis}]{tsirvoulis2018reconstructing}
Tsirvoulis, G., Morbidelli, A., Delbo, M., \& Tsiganis, K. 2018, Icarus, 304,
  14

\bibitem[{Usui {et~al.}(2011)Usui, Kuroda, M{\"u}ller, Hasegawa, Ishiguro,
  Ootsubo, Ishihara, Kataza, Takita, Oyabu, {et~al.}}]{usui2011asteroid}
Usui, F., Kuroda, D., M{\"u}ller, T.~G., {et~al.} 2011, Publications of the
  Astronomical Society of Japan, 63, 1117

\bibitem[{{{\v{D}}urech} {et~al.}(2011){{\v{D}}urech}, {Kaasalainen}, {Herald},
  {Dunham}, {Timerson}, {Hanu{\v{s}}}, {Frappa}, {Talbot}, {Hayamizu},
  {Warner}, {Pilcher}, \& {Gal{\'a}d}}]{Durech2011Icar..214..652D}
{{\v{D}}urech}, J., {Kaasalainen}, M., {Herald}, D., {et~al.} 2011, \icarus,
  214, 652

\bibitem[{Vernazza {et~al.}(2021)Vernazza, Ferrais, Jorda, Hanu{\v{s}}, Carry,
  Marsset, Bro{\v{z}}, F{\'e}tick, Viikinkoski, Marchis,
  {et~al.}}]{vernazza2021vlt}
Vernazza, P., Ferrais, M., Jorda, L., {et~al.} 2021, Astronomy \& Astrophysics,
  654, A56

\bibitem[{Vokrouhlicky {et~al.}(2015)Vokrouhlicky, Bottke, Chesley, Scheeres,
  \& Statler}]{vokrouhlicky2015yarkovsky}
Vokrouhlicky, D., Bottke, W.~F., Chesley, S.~R., Scheeres, D.~J., \& Statler,
  T.~S. 2015, arXiv preprint arXiv:1502.01249

\bibitem[{Vokrouhlick{\'{y}} {et~al.}(2006)Vokrouhlick{\'{y}}, Bro{\v{z}},
  Bottke, Nesvorn{\'{y}}, \& Morbidelli}]{Vokrouhlicky2006}
Vokrouhlick{\'{y}}, D., Bro{\v{z}}, M., Bottke, W.~F., Nesvorn{\'{y}}, D., \&
  Morbidelli, A. 2006, Icarus, 182, 118

\bibitem[{Walsh {et~al.}(2013)Walsh, Delb{\'{o}}, Bottke, Vokrouhlick{\'{y}},
  \& Lauretta}]{Walsh2013}
Walsh, K.~J., Delb{\'{o}}, M., Bottke, W.~F., Vokrouhlick{\'{y}}, D., \&
  Lauretta, D.~S. 2013, Icarus, 225, 283

\bibitem[{Zappala {et~al.}(1990)Zappala, Cellino, Farinella, \&
  Knezevic}]{zappala1990asteroid}
Zappala, V., Cellino, A., Farinella, P., \& Knezevic, Z. 1990, The Astronomical
  Journal, 100, 2030

\end{thebibliography}

\begin{appendix}
    \section{Membership of known families}\label{appendix:knownFamMembhersip}

\begin{figure*}
        \centering
        \includegraphics[width=\linewidth]{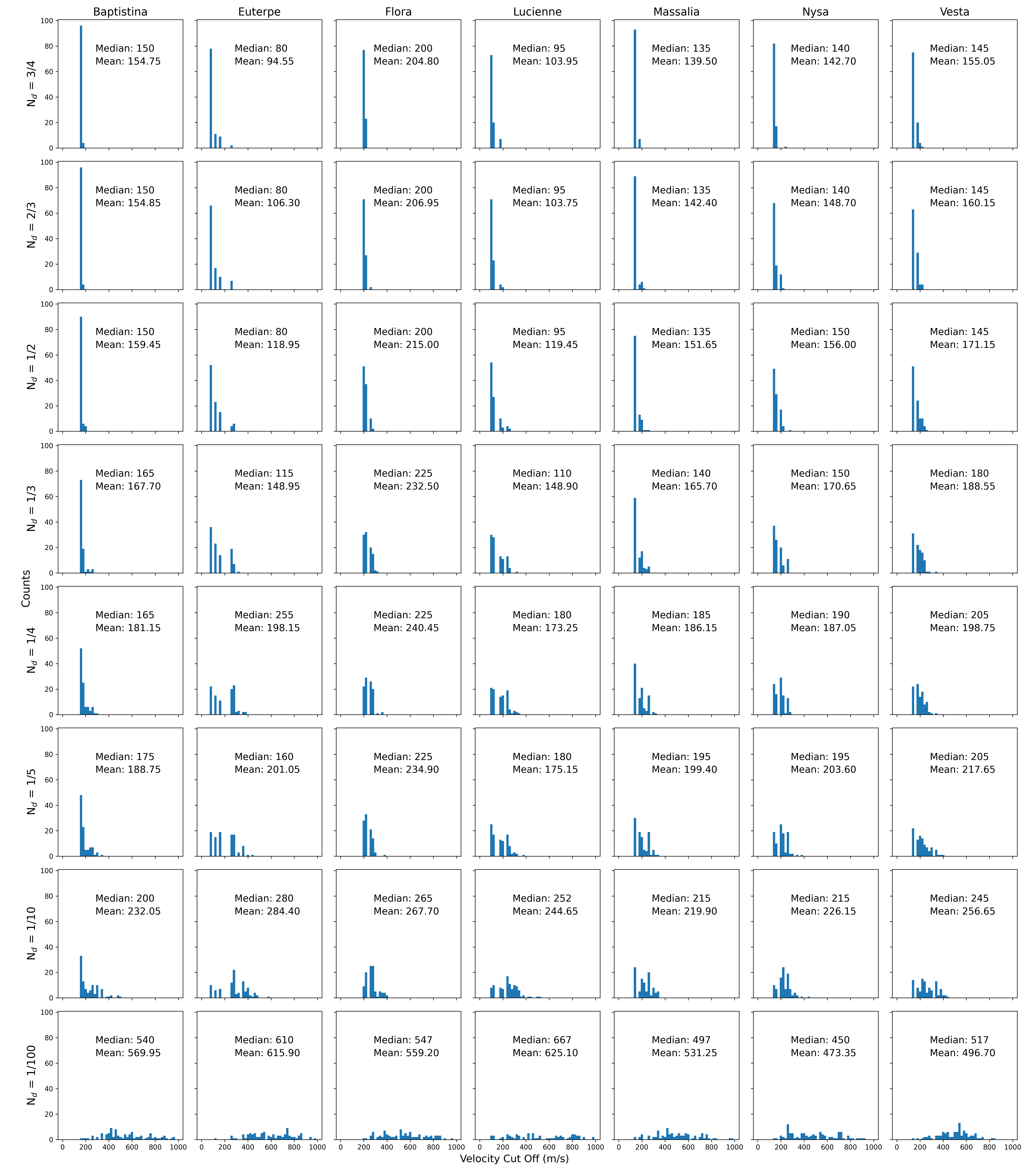}
        \caption{Velocity cut-off as a function of the density of the artificial background}
        \label{fig:velocityCutOffHistogram}
    \end{figure*}

    First, while choosing $v_c$ we know the results are going to be sensitive to the number density of particles in the proper element space of the reference synthetic data set. This synthetic data set from \cite{Deienno2021} may be an overestimate. Therefore, we created nine different samplings, beginning from all 18,191 bodies provided in the dataset, and narrowed this value to 182, which is certainly unrealistically few. As explained in section~\ref{sec:V-shape-constrained-hcm}, for each number density below 1, we randomly remove bodies 100 different ways. Then, for each, we apply the HCM for each central body of the families listed in Table~\ref{tab:hcm-simulated-bg}. The distributions of the found velocities are reported in Fig.~\ref{fig:velocityCutOffHistogram}. As expected, the chosen $v_c$ increases as the density decreases. This is always true for the mean and generally true for the median. As seen in the last column, where the density if 1/100, the values are rather spread and no longer consistent with the higher-density background.

\begin{figure*}
        \includegraphics[width=\linewidth]{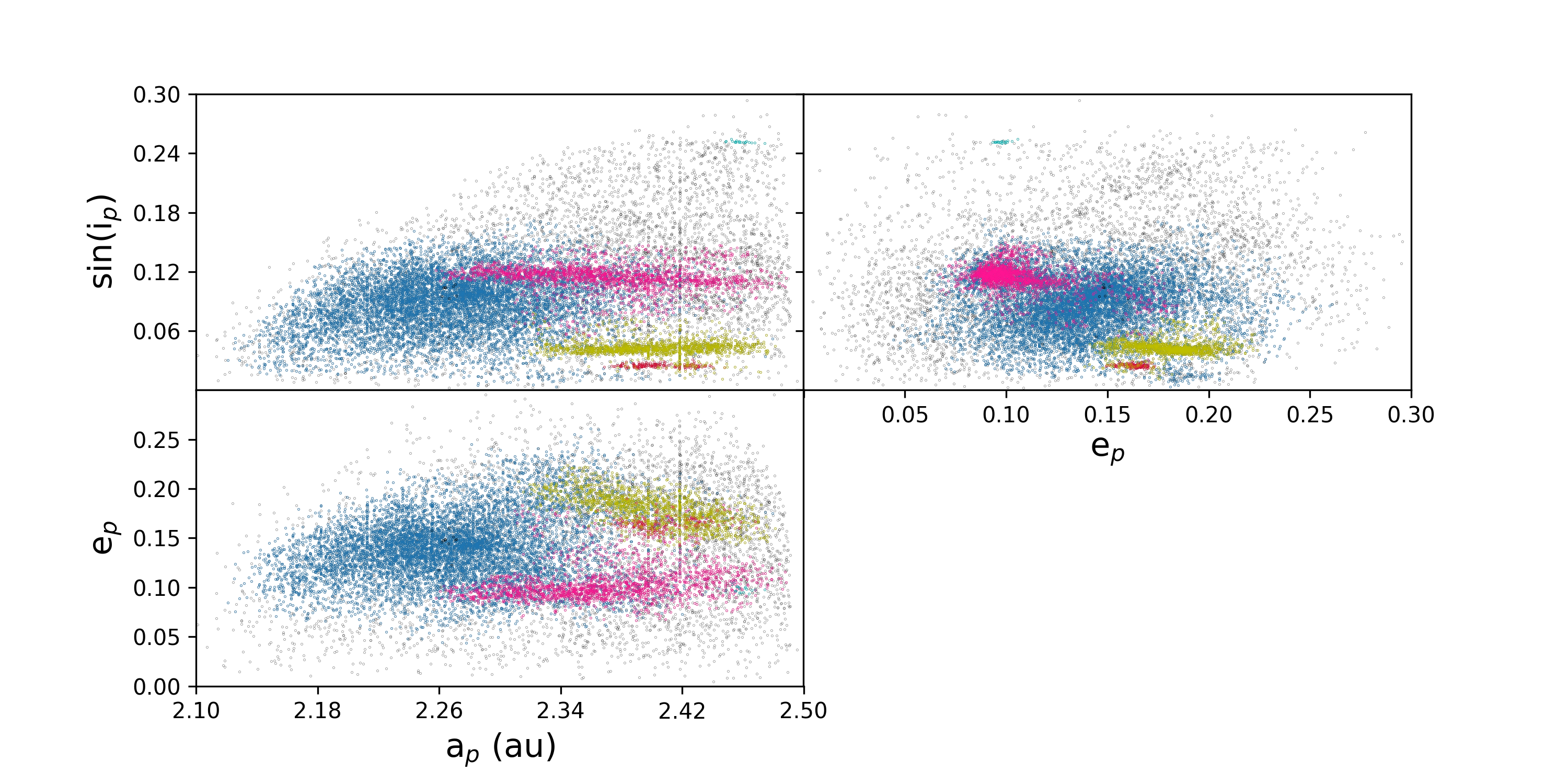}
        \includegraphics[width=\linewidth]{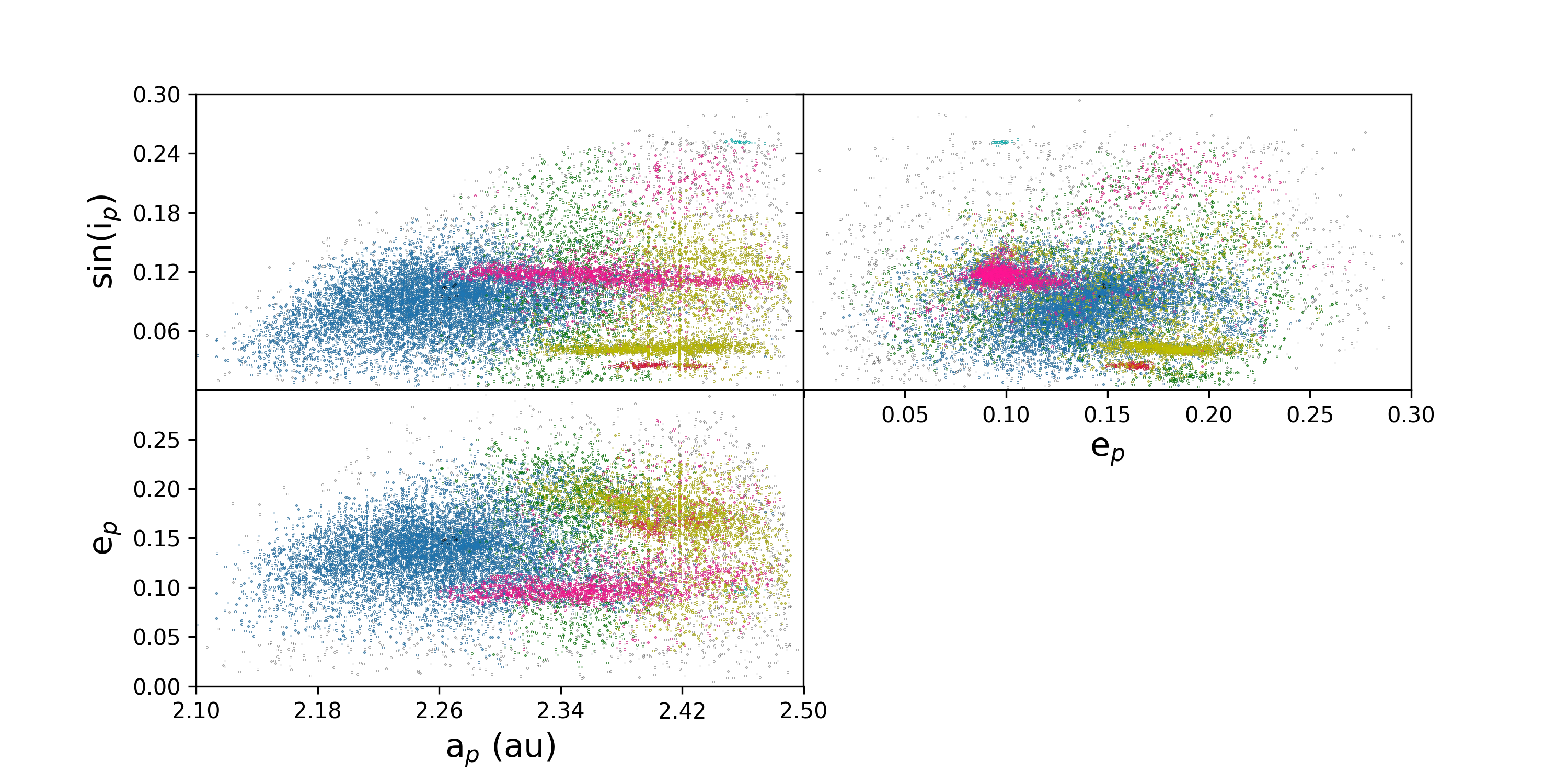}        
        \caption{Membership of each family. The Flora--Baptistina complex is shown in blue, Euterpe in green, Vesta in red, Nysa in yellow, Massalia in purple, and Lucienne in cyan. Background objects are in gray. The top panel uses the $v_c$ values defined using a synthetic background of 3/4 from Table~\ref{tab:hcm-simulated-bg}, while the bottom panel uses 1/10.}\label{fig:coloredFamilies}
    \end{figure*}

    The main goal of our experiment is to remove all bodies that are members of collisional families. We were not concerned with providing precise membership. For instance, in our method, a body can be assigned to multiple families, as highlighted by the part of Vesta assigned captured in Baptistina in Fig.~\ref{fig:baptistina-vshape}. Despite this, we were interested in providing a membership list anyway. To do so, we created a matrix where each row was an asteroid, each column was a family, and each element was the $v_c$ at which the body is assigned to the corresponding family. Then, we subtracted each column by the $v_c$ at which the family began clustering the second body. After this, we assigned each body to the family based on the minimum value. For instance, if body X is assigned to Flora at 50 $\textrm{m s}^{-1}$ and Vesta at 40 $\textrm{m s}^{-1}$, the body in the end will be assigned to Vesta. These results are displayed in Fig.~\ref{fig:coloredFamilies}. 

    Originally, we did not normalize to the velocity at which the second body is assigned to the cluster. However, membership was not well assigned, especially in the case of Nysa and Massalia, which became confused with one another. Shifting by the second body gathers more of the essence of the cluster's center rather than the central body that takes the families name. 

    One suboptimal result is that this method is not able to parse Baptistina from Flora. Only four bodies are assigned to Baptistina. Thus, we believe separating these families beyond the cores is more involved and future work is needed to tackle this problem.

    \section{Membership of planetesimals and ancient family members}\label{appendix:ancient-primordial-members}
\begin{figure*}
        \includegraphics[width=\linewidth]{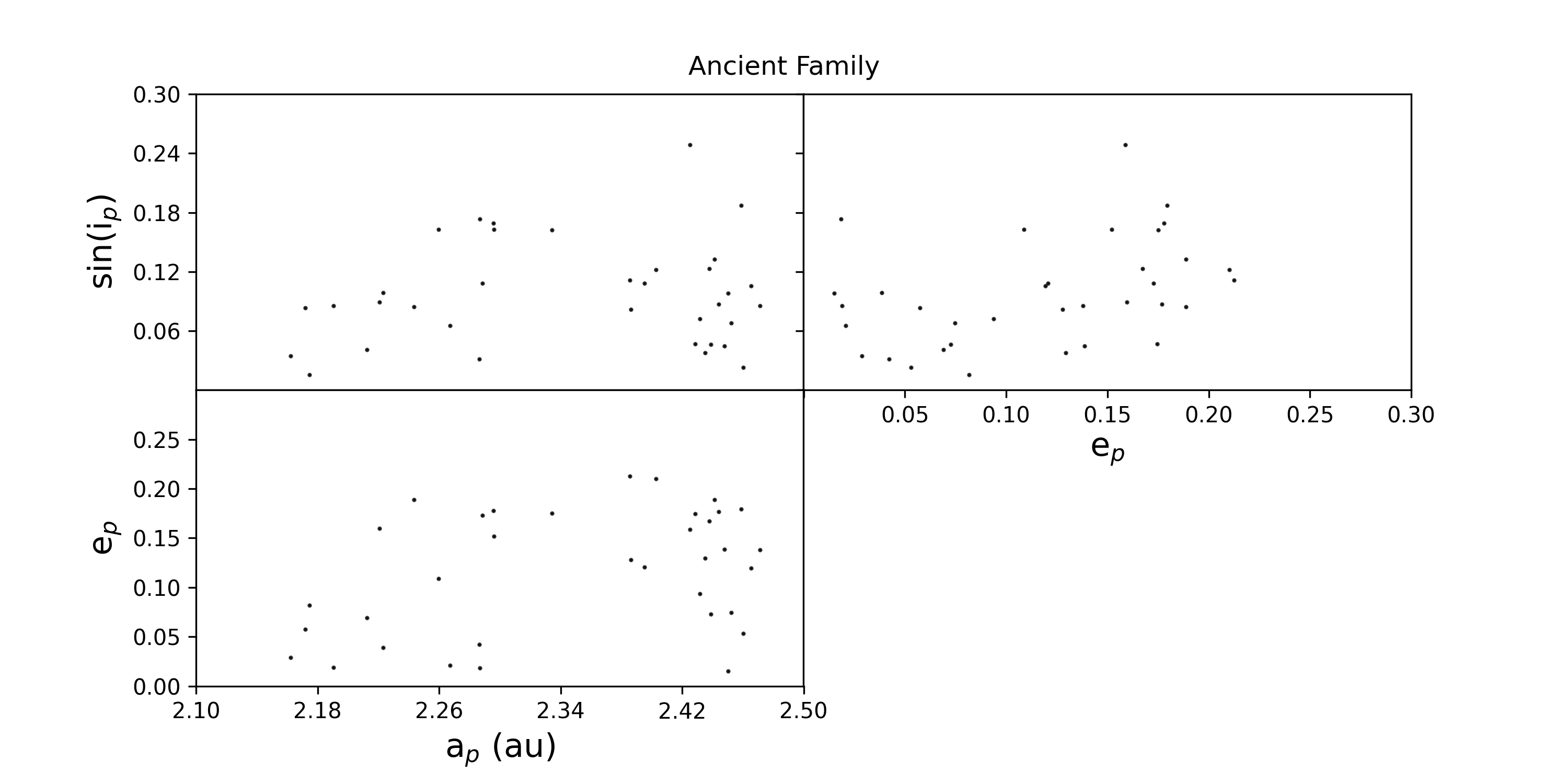}
        \caption{Proper elements of the ancient family members reported in Table~\ref{tab:ancient-family}.}\label{fig:proper-elements-ancient-family}
    \end{figure*}

    The lists of the ancient family, planetesimals, and the bodies that could belong to either are presented in Tables~\ref{tab:ancient-family},~\ref{tab:hcm-simulated-bg}, and ~\ref{tab:eitherOR}, respectively. Additionally, these objects are labeled with red, magenta, and yellow markers in the left panel of Fig.~\ref{fig:ancient-family-detection}. Lastly, these are the bodies that were assigned to each group at a 1$\sigma$ significance level, as described in section~\ref{sec:vsearch}. The distribution of the ancient family members is shown in Fig.~\ref{fig:proper-elements-ancient-family}.
    
\begin{table}
    \centering
    \caption{List of objects within the lobes of the V-shape used to identify the ancient family.}
    \label{tab:ancient-family}
    \begin{tabular}{|l|l|l|}
    \hline
    (30) Urania & (470) Kilia & (1365) Henyey\\
    (60) Echo & (584) Semiramis & (1375) Alfreda\\
    (115) Thyra & (753) Tiflis & (1419) Danzig\\
    (161) Athor & (822) Lalage & (1594) Danjon\\
    (169) Zelia & (889) Erynia & (1629) Pecker\\
    (172) Baucis & (896) Sphinx & (1643) Brown\\
    (186) Celuta & (939) Isberga & (1689) Floris-Jan\\
    (219) Thusnelda & (1063) Aquilegia & (1797) Schaumasse\\
    (234) Barbara & (1078) Mentha & (1988) Delores\\
    (287) Nephthys & (1117) Reginita & (2013) Tucapel\\
    (299) Thora & (1137) Raissa & (2159) Kukkamaki\\
    (306) Unitas & (1147) Stavropolis & (2286) Fesenkov\\
    (337) Devosa & (1155) Aenna & (3385) Bronnina\\
    (432) Pythia & (1224) Fantasia & (5676) Voltaire\\
    \hline 

    \end{tabular}
    \tablefoot{These bodies are those of Fig.~\ref{fig:ancient-family-detection}. Mass and diameter estimates are available for: 30,  60, 115, 172, 287, 337, 432, 584, and 939 from \cite{vernazza2021vlt}, which were used to calculate a density and in turn obtain one of the age estimations.
    }
\end{table}

\begin{table}
    \centering
    \caption{List of nominal planetesimals. }
    \label{tab:primordial-objects}    
    \begin{tabular}{|l|l|l|}
        \hline
        (6) Hebe & (79) Eurynome & (192) Nausikaa\\
        (7) Iris & (80) Sappho & (198) Ampella\\
        (9) Metis & (118) Peitho & (317) Roxane\\
        (11) Parthenope & (126) Velleda & (364) Isara\\
        (12) Victoria & (131) Vala & (376) Geometria\\
        (17) Thetis & (135) Hertha & (556) Phyllis\\
        (18) Melpomene & (136) Austria & (722) Frieda\\
        (21) Lutetia & (138) Tolosa & (813) Baumeia\\
        (40) Harmonia & (149) Medusa & (857) Glasenappia\\
        (42) Isis & (178) Belisana & (1182) Ilona\\
        (63) Ausonia & (189) Phthia & (2616) Lesya\\
         &  & (2675) Tolkien\\
    \hline
    \end{tabular}
    \tablefoot{These asteroids lie below the V-shape in Fig.~\ref{fig:ancient-family-detection}.}
\end{table}

\begin{table}
    \centering
    \caption{Asteroids that are either planetesimals or ancient family members}
    \label{tab:eitherOR}
    \begin{tabular}{|l|l|l|}
    \hline
    (230) Athamantis & (525) Adelaide & (548) Kressida\\
    (749) Malzovia & (1396) Outeniqua & \\
    \hline
    \end{tabular}
    \tablefoot{These objects are displayed as yellow squares with black edges in Fig.~\ref{fig:ancient-family-detection}.}
\end{table}

\begin{table*}[]
    \centering
    \caption{Complete list of planetesimals used to create the SFD in the bottom panel of Fig.~\ref{fig:SDFs}. These objects come from Table~\ref{tab:primordial-objects} as well as the planetesimals from \citet{Delbo2017} and \citet{Delbo2019}.} \label{table:allPlanetesimals}

    \begin{tabular}{|l|l|l|l|l|l|}
    \hline
    (4) Vesta       & (20) Massalia & (80) Sappho   & (178) Belisana   & (336) Lacadiera & (689) Zita        \\
    (6) Hebe        & (21) Lutetia  & (118) Peitho  & (189) Phthia     & (337) Devosa    & (722) Frieda      \\
    (7) Iris        & (27) Euterpe  & (126) Velleda & (192) Nausikaa   & (345) Tercidina & (813) Baumeia     \\
    (8) Flora       & (40) Harmonia & (131) Vala    & (198) Ampella    & (364) Isara     & (857) Glasenappia \\
    (9) Metis       & (42) Isis     & (135) Hertha  & (207) Hedda      & (376) Geometria & (1182) Ilona      \\
    (11) Parthenope & (51) Nemausa  & (136) Austria & (261) Prymno     & (435) Ella      & (1892) Lucienne   \\
    (12) Victoria   & (63) Ausonia  & (138) Tolosa  & (298) Baptistina & (556) Phyllis   & (2616) Lesya      \\
    (17) Thetis     & (72) Feronia  & (149) Medusa  & (317) Roxane     & (572) Rebekka   & (2675) Tolkien    \\
    (18) Melpomene  & (79) Eurynome & (161) Athor   & (326) Tamara     & (654) Zelinda   & (13977) Frisch    \\
    \hline
    \end{tabular}
\end{table*}

\section{Planetesimal counts compensation}\label{appendix:vcomp}

\begin{figure}
    \includegraphics[width=\linewidth]{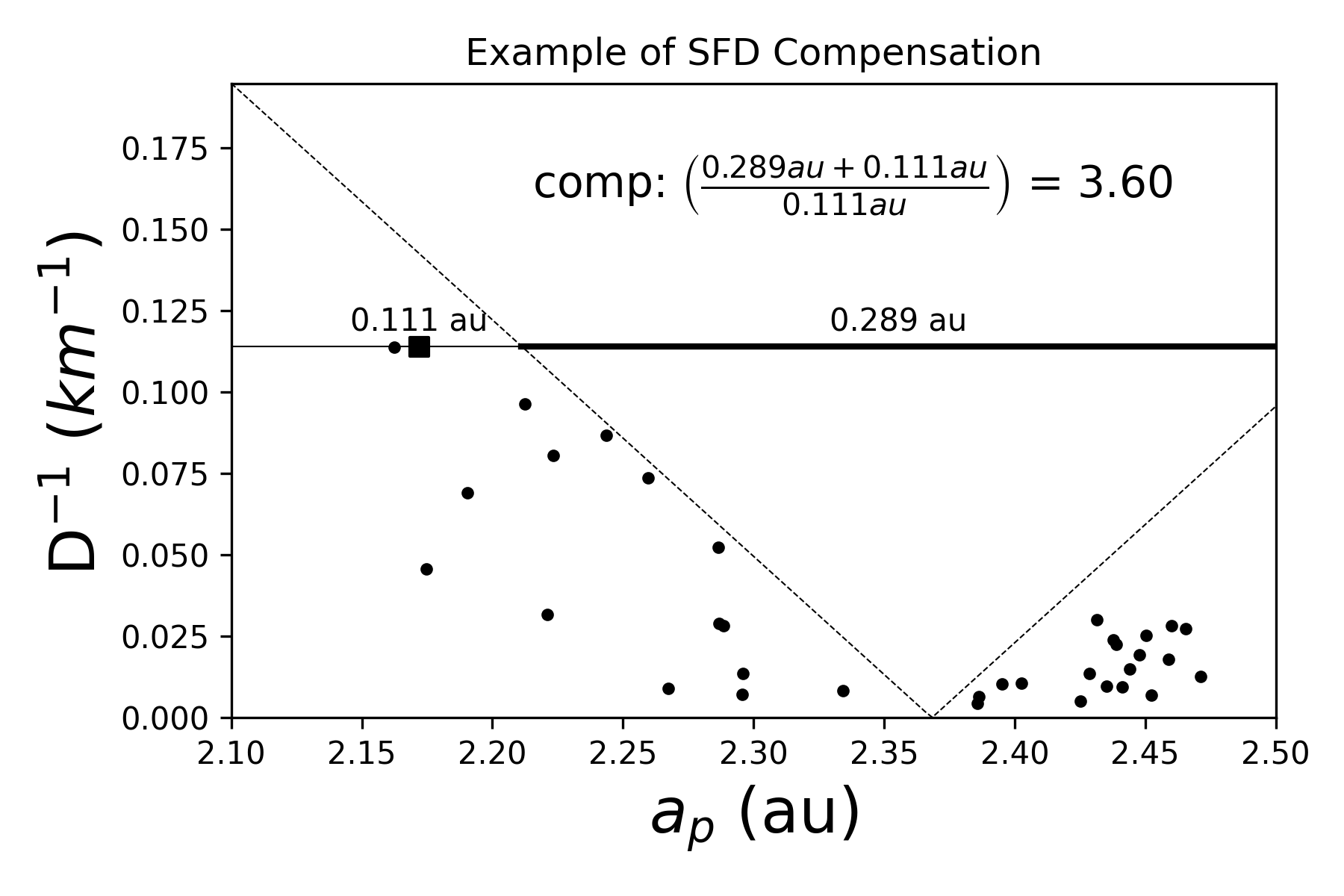}

    \includegraphics[width=\linewidth]{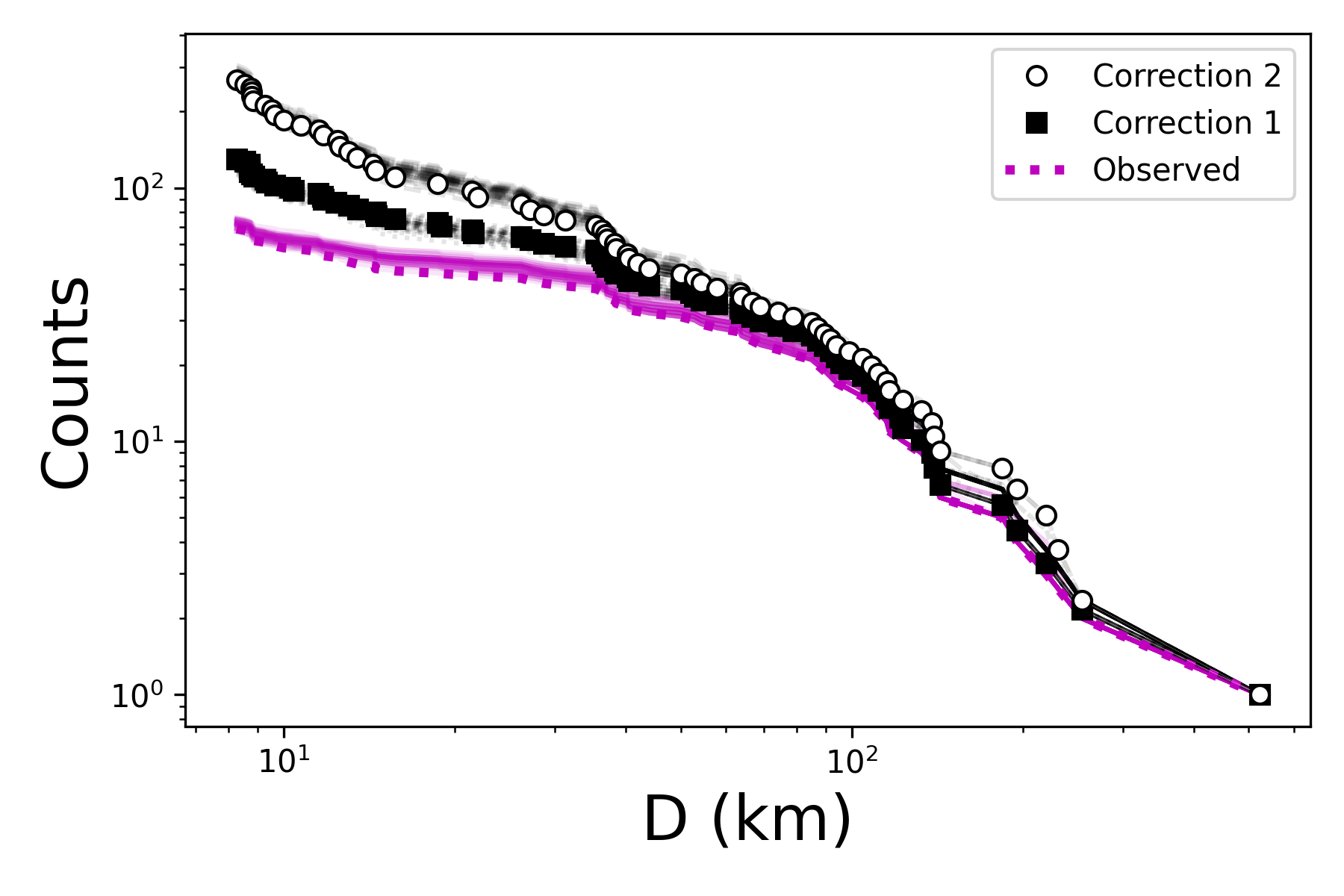}
    \caption[]{\textit{Top:} Demonstration of compensating the number of planetesimal counts missed due to the ancient family covering a large portion of the search area. Here, we show the V-shape corresponding to the newly detected family in this work. The black points are the identified planetesimals from Table~\ref{tab:primordial-objects}. The black square is an example case, whose count is increased in proportion to the amount of search area that was obscured by the family's V-shape. Here, the counts will increase from one to three, with a 60\%\ chance of being increased to four. See the text for more details. \textit{Bottom:} Compensation was applied 50 times with Monte Carlo-style sampling and all iterations are over-plotted with highly transparent lines.  }
    \label{fig:vcomp}
\end{figure}

As shown in Fig.~\ref{fig:vcomp}, the identification method for finding the planetesimals has a bias. Namely, no bodies can be found within V-shapes, yet \emph{a priori} there is no reason that planetesimals cannot exist within these areas. Though we cannot identify which of the asteroids within the V are likely planetesimals with the dynamical and size information alone, we can estimate how many we could expect to exist within the V-shapes. In turn, we can increase the number of counts on a per asteroid basis by taking the ratio of the total linear length, in semi-major axis, divided by the available search length, at a given asteroid diameter. The available search length exists outside the V. The total search length is always equal to 0.4~au, which is the limit of the inner main belt between 2.1 and 2.5~au. This, of course, is also equal to the sum of lengths interior and exterior to the V. For example, for the asteroid in Fig.~\ref{fig:vcomp}, this equates to 3.6. 

If computing the SFD directly, we could simply increase the weight of each asteroid in the cumulative sum from 1 to the compensation value. However, the dynamical and collisional loss models requires that we work with equally weighted asteroids. To circumvent this problem, we generate a number of asteroids corresponding to the suggested compensation on a per-asteroid basis in a Monte Carlo fashion and append these generated asteroids to the total number of asteroids in Table~\ref{table:allPlanetesimals}. To specify, and in the case of the asteroid in Fig.~\ref{fig:vcomp},  we append the list by two more asteroids, such that our count for this diameter reaches three. Next, we generate a random number from a uniform probability density distribution between 0 and 1. If this number is less than 0.6, then we add a third asteroid to the list. For each of these new bodies, in order to avoid duplicates, we give them new diameters by Gaussian sampling the diameter of the corresponding planetesimal given its uncertainty. This process is repeated for all of the asteroids. We also perform this random sampling 50 times. In each iteration, we typically add between 12 and 25 bodies. In general, the large objects have compensation values on the order of 1.03, which leads to small probabilities of adding large bodies; most of the new bodies have small diameters.

On a separate note, the V-compensation was applied according to the  V-shape of the ancient family that was used to isolate the planetesimal. For instance, all of the planetesimals from Table~\ref{tab:primordial-objects} use the V shape of the family identified in this work, whereas the planetesimals identified in \citet{Delbo2017} and \citet{Delbo2019} used the V-shape of the family used in those respective studies. For the parent bodies of asteroid families that were added, such as Flora, Vesta, Euterpe, and so on, no V-compensation was performed because they were not identified with this method.

Lastly, we plot all 50 iterations of the resulting SFD in the bottom panel of Fig.~\ref{fig:vcomp}. Interestingly, the variance of possible SFDs resulting from considering the Monte Carlo-style generation of new asteroids while considering the uncertainties of the diameters is much less than the variance from modeling the uncertainty used to correct the dynamical loss. For this reason, in the main plot of Fig.~\ref{fig:SDFs}, where we fit the power-law slopes, we only present one iteration for clarity. 

\end{appendix}

\end{document}